\documentclass[useAMS,usenatbib]{mn2e}
 \usepackage{natbib}
\usepackage{graphicx}
\pdfoutput=1
\title[Stellar population of NGC~628]{PPAK Wide field Integral Field Spectroscopy of NGC~628 III.
Stellar population properties}
\author[S\'anchez-Bl\'azquez et al.]{P. S\'anchez-Bl\'azquez$^{1}$\thanks{E-mail:p.sanchezblazquez@uam.es}, F. Rosales-Ortega$^{1,2}$, 
A. Diaz$^{1}$; S.F.~S\'anchez$^{3,4}$\\
$^{1}$Departamento de Fisica Teorica, Universidad Aut\'onoma de Madrid, Cantoblanco, 28049, Madrid, Spain\\
$^{2}$Instituto Nacional de Astrof{\'i}sica, {\'O}ptica y Electr{\'o}nica, Luis E. Erro 1, 72840 Tonantzintla, Puebla, Mexico\\
$^{3}$Instituto de Astrof\'{\i}sica de Andaluci\'{\i}a (CSIC), Glorieta de la Astronom\'{\i}a s/n, E18080, Granada, Spain\\
$^{4}$Centro Astron\'{o}mico Hispano Alem\'an de Calar Alto (CSIC-MPIA), E4004, Almer\'{\i}a, Spain}
\begin{document}

\date{Accepted 1988 December 15. Received 1988 December 14; in original form 1988 October 11}

\pagerange{\pageref{firstpage}--\pageref{lastpage}} \pubyear{2002}

\maketitle

\label{firstpage}

\begin{abstract}
We present a stellar population analysis of the nearby, face-on, SA(s)c galaxy, NGC~628, which is 
part of the PPAK IFS Nearby Galaxies Survey (PINGS). The data cover a field of view of 
$\sim$ 6 arcmin in diameter with a sampling of $\sim$2.7~arcsec per spectrum and a wavelength
range (3700-7000\AA).  We apply spectral inversion methods to derive 2-dimensional maps of 
star formation histories and chemical enrichment.
We present  maps of the mean (luminosity- and mass-weighted) age and metallicity that 
reveal the presence of structures such as a nuclear ring, previously seen in molecular gas. 
The disk is dominated in mass by an old stellar component at all radii sampled by our data,
while the percentage of young stars increase with radius.
The mean stellar age and metallicity profiles have a two defined regions, an inner one
with flatter gradients (even slightly positive) and an external ones with a negative, steeper one, separated at $\sim$60 arcsec.
This break in the profiles is more prominent in the old stellar component.
The young component shows a metallicity gradient that is very similar to that of the gas, 
and that is flatter in the whole disc. 
The agreement between the metallicity gradient of the young stars and the gas, and the recovery of the 
measured colours from our derived star formation histories validate the techniques to recover 
the age-metallicity and the star formation histories in disc galaxies from integrated spectra.
We speculate about the possible origin of the break and conclude that the most likely scenario
is that we are seeing, in the center of NGC~628, a dissolving bar, as predicted in some numerical simulations.
\end{abstract}


\section{Introduction}

Quantifying the  star formation histories of galaxies constitutes one of the major unsolved
issues towards a complete understanding of galaxy formation. Although this task is difficult, 
analyses of stellar populations  constitute a step forward toward the achievement of this goal.
In the majority of cases, these studies have to be made using integrated spectra or colours, as we can 
only resolve stars in a limited number of galaxies. 
While studies of stellar populations for early-type galaxies are abundant in the literature
(e.g., Trager et al. 2000\nocite{Trageretal.2000}; Kuntschner 2000\nocite{Kuntschner2000}, 
Thomas et al. 2005\nocite{Thomasetal.2005}; S\'anchez-Bl\'azquez
et al. 2006abc\nocite{Sanchez-Blazquezetal2006a}\nocite{Sanchez-Blazquezetal2006b}; 
Smith, Lucy \& Hudson 2007\nocite{Smithetal.2007}, among may others), 
these studies are much more sparse for disc galaxies. 
Until very recently, the study of the stellar component in disk galaxies outside the local 
group was restricted to broadband photometric data (Bell \& de Jong 2000\nocite{Bell&deJong2000}; 
MacArthur et al. 2004\nocite{MacArthuretal.2004};
Pohlen \& Trujillo 2006\nocite{Pohlen&Trujillo2006}; Mu\~noz-Mateos et al.
2007\nocite{Munoz-Mateosetal.2007}; Roediger et al. 2012\nocite{Roedigeretal.2012}). 
These studies find that  disc galaxies tend to be bluer in the external parts,
a trend that is usually interpreted in terms of stellar population gradients as galaxies tend to be  
younger and more metal poor in the external parts. However, there are large discrepancies 
in the magnitude of the stellar population gradients derived by different authors due, 
partially, to the difficulty in disentangling the effects of the age, metallicity and 
dust extinction by using only colours. Spectroscopic studies may help to alleviate 
the associated degeneracies but the low surface brightness of the
disc region and the nebular emission lines filling some of the most important age-diagnostic 
absorption features make these studies very difficult. In the last few years,  however, the 
advent of new techniques to separate the contributions of  stellar and gaseous components 
(see,e.g., Sarzi et al.\ 2005)\nocite{Sarzietal.2005} and  large aperture telescopes have allowed the obtention of 
enlightening results in this field  (see, e.g.,  Yoachim \& Dalcanton 2008\nocite{Yoachim&Dalcanton2008}; 
MacArthur, Gonz\'alez \& Courteau 2009\nocite{MacArthurGonzalez&Courteau2009}; 
P\'erez et al. 2011; S\'anchez-Bl\'azquez et al. 2011\nocite{Sanchez-Blazquezetal.2011}).

All the studies quoted above used long-slit spectroscopy,
focusing along a predefined spatial axis. 
Disc galaxies, however,  have a complex geometry, with many 
structural components, such as bulges, bars, rings, etc, which makes the interpretation of the
long-slit data a difficult task. Integral Field spectroscopy (IFS, hereafter)
allows the study of the different galaxy components mentioned above separately.
IFS data cubes provide, for each pixel of the galaxy, a spectrum from which absorption and 
emission lines can be extracted, allowing to decipher the star formation history and the 
physical properties of the interstellar medium in galaxies.
Also, IFS offers a unique opportunity to improve the signal to noise 
ratio of the data by averaging many more spectra in the  external parts of the galaxy and makes 
easier the masking or the extraction  of the HII regions in the spatial map. In the 
last few years, there have been some studies of the stellar population properties of 
disc galaxies from 
IFS spectroscopy (Ganda et al.\ 2007\nocite{Gandaetal.2007}; S\'anchez et al. 2011\nocite{Sanchezetal.2011}; Yoachim et al. 2009,
2012\nocite{Yoachimetal.2009}\nocite{Yoachimetal.2012}),      
probing the potential of this technique. In fact, many ongoing and planned surveys are devoted to the observations 
of nearby galaxies using Integral Field Units (e.g., CALIFA (S\'anchez et al. 2011)\nocite{Sanchezetal.2011}; 
VENGA (Blanc et al. 2010)\nocite{Blancetal.2010}; 
SAMI (Croom et al. 2012)\nocite{Croometal.2012}; MANGA (K. Bundy)).

A further complication in the study of stellar population in disc galaxies is that, surely, the stars did not form 
in a single burst of a given age and metallicity as usually assumed in studies of early-type galaxies. Therefore, the 
comparison of spectral characteristics with single stellar populations (SSP) models 
leads to information which is difficult
to interpret. Fortunately,  in the last decade there has been a tremendous improvement in the stellar population analysis techniques,     
associated with the release  of well-calibrated stellar libraries (e.g., STELIB, LeBorgne et al. 2003\nocite{LeBorgneetal.2003};                                                      
MILES, S\'anchez-Bl\'azquez et al. 2006\nocite{Sanchez-Blazquezetal2006d}; 
Indo-US (Valdes et al.\ 2004\nocite{Valdesetal.2004}); CaT (Cenarro et al. 2001ab)\nocite{Cenarroetal.2001a}
\nocite{Cenarroetal.2001b} that                                                    
have allowed the construction of stellar population models that predict the whole synthetic spectrum                                                        
of a population with a given age and metallicity (e.g., Vazdekis 1999\nocite{Vazdekis1999}; 
Vazdekis et al. 2003\nocite{Vazdekisetal.2003}, 2010;
Bruzual \& Charlot 2003\nocite{Bruzual&Charlot2003}, Conroy et al. 2009,
2010ab\nocite{Conroyetal2009}\nocite{Conroy2010a}\nocite{Conroy2010b}; Maraston \& Str\"ombäck 201\nocite{MarastonStromback2011}1). This has                                                        
driven the development of inversion algorithms to extract information about  the  star formation
histories and the metallicity evolution from integrated spectra (e.g., {\tt MOPED}: Heavens et al. 2000\nocite{Heavensetal.2000}, 
Panter et al. 2003\nocite{Panteretal.2003};                                
{\tt STARLIGHT}: Cid Fernandes et al. 2005\nocite{CidFernandesetal.2005}, 
Moultaka et al. 2005\nocite{Moultaka2005}; {\tt STECKMAP}: Ocvirk et al.\ 2006ab\nocite{Ocvirketal.2006a}\nocite{Ocvirketal.2006b}; {\tt VESPA}: Tojeiro
et al. 2007\nocite{Tojeiroetal.2007}; {\tt FIT3D}: S\'anchez et al. (2008); {\tt ULySS}: Koleva et al. 2009\nocite{Koleva2009}).                                                                     
The combination of these two developments make the study of the
stellar population properties in disc galaxies very timely.                

In the present paper we derive the stellar population properties and
star formation histories of  NGC~628 in a 3-dimensional fashion,
using several techniques.
NGC~628 (M74) is an extensively studied isolated grand-design SA(s)c (Corwin et al. 1994)
\footnote{There is abundant photometry in different bands (Holmberg 1975\nocite{Holmberg1975}; Boroson 1981\nocite{Boroson1981};
Shostak \& van der Kruit 1984\nocite{Shostak&vanderKruit1984};Natali, Pedichini \& Righini 1992\nocite{Natalietal.1992};
Hoopes et~al.~2001\nocite{Hoopesetal.2001}; Kennicutt et al. 2008\nocite{Kennicuttetal.2008}; 
Gil de Paz et al. 2007a\nocite{GildePazetal.2007}).} 
spiral galaxy at a distance of  9.3 Mpc in the constellation of Pisces. 
It does not show any evidence of having had encounters with satellites or other galaxies                                         
in the last $10^9$ yr (Kamphuis \& Briggs 1992\nocite{Kamphuis&Briggs1992}).                                               
Regarding its structure, NGC~628   shows an  outer HI ring at around 12 arcmin from the 
nucleus (Roberts 1962\nocite{Roberts1962};                 
Briggs et al. 1980\nocite{Briggsetal.1980}) that                                                                           
seems to be due to the interaction with two large high-velocity clouds accreting onto the outer                            
parts of the disc (Kamphuis \& Briggs 1992\nocite{Kamphuis&Briggs1992};                                                    
L\'opez-Corredoira et al. 2002\nocite{Lopez-Corredoiraetal.2002}; Beckman et~al.\ 2003\nocite{Beckmanetal.2003}).          
It also has a inner rapidly rotating disc-like structure (Baigle et al. 2006), a circumnuclear   ring                      
of  star formation at $\sim$2 kpc from the center (discovered in $^{12}$CO J=1-0 sub-mm, Wakker \& Adler                   
1995\nocite{Wakker&Adler1995}, and 2.3 $\mu$m CO absorption, James \& Seigar 1999\nocite{James&Seigar1999})                
and a nuclear, nested bar on a                               
$\sim$100 pc scale (Laine et al. 2002)\nocite{2002ApJ...567...97L}.                                                                                    
An oval structure has also being detected at a radii around 50$" \sim 2.3$~kpc (Seigar 2002\nocite{Seigar2002}).
All these characteristics indicate that secular evolution processes are taking place in this galaxy                        
(see Fathi et al. 2007\nocite{Fathietal.2007}).      

This paper can  be considered as a pilot study that allows us to
explore the techniques that will be applied to the study of
a larger sample of galaxies from the CALIFA survey (S\'anchez et al.\ 2012\nocite{Sanchezetal.2012}).
 The study complements and expands the ones 
presented in S\'anchez et al. (2011, paper I) and Rosales-Ortega et al.~(2011, paper II) for the same galaxy.
NGC~628 is the largest object within the PPAK Integral Field 
Spectroscopy Nearby Galaxies Survey: PINGS (Rosales-Ortega et al.\
2010\nocite{Rosales-Ortegaetal2010}). PINGS IFS survey of nearby ($<$100 Mpc) well-resolved spiral
galaxies is specially designed to obtain complete maps of the
emission-line abundances, stellar populations, 
and reddening using an IFS mosaicking imaging, which takes advantage 
of what is currently one of the world$'$s widest 
field-of-view (FOV) integral field units.

Section~\ref{sec_obs} briefly describes the observations,
Sec.~\ref{sec:analysis} the analysis performed to derive the stellar
population properties and, in Sec.~\ref{sec_results} we present our 
results. Section~\ref{sec:comparison} tests
the robustness of the solution comparing with the analyses performed
with different techniques and by different authors. 
In Sec.~\ref{sec_discussion} we provide a brief discussion and 
Sec.~\ref{sec_conclussion} summarizes our conclusions. 

\section{Observations}
\label{sec_obs}
The observations analysed here are part of the PPAK IFS Nearby Galaxies Survey 
(PINGS, Rosales-Ortega  et al.\ 2010\nocite{Rosales-Ortegaetal2010}).
The PPAK fibre bundle consists of 382 fibers of 2.7$"$ diameter each. Of these 382 fibers,
331 are concentrated in a single hexagonal bundle covering a field-of-view of 72$\times$63$"$,
with a filling factor of $\sim$65\%. The sky background is sampled by 36 additional fibers in 
6 bundles of 6 fibers each, distributed along a circle of $\sim$90 arcsec from the center.
Due to the large size of NGC~628 compared to the field-of-view of the instrument,
a mosaicking scheme was adopted.  The central position was observed
in dithering mode to gain spatial resolution, while the remaining 
positions
were observed without dithering due to the large size of the mosaic.

\begin{figure*}
\centering
\resizebox{0.5\textwidth}{!}{\includegraphics[angle=90]{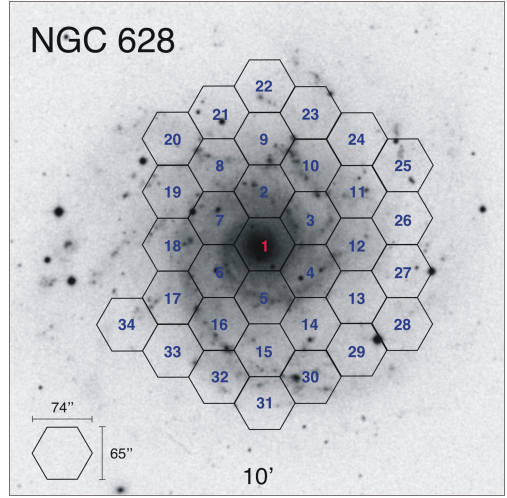}}
\caption{$B$-band Digital Sky Survey image of NGC\,628. The mosaic of the PPAK pointings is shown 
as overlaid hexagons indicating the
field-of-view of the central fiber-bundle. The numbers correspond to the identification number of each pointing. The image is
10'$\times$10$'$ and it is displayed in top-north, left-east standard configuration.\label{mosaico_n628}}
\end{figure*}

The observations for this galaxy extended over a period of three years in
different stages, with a total of six observing nights.
The spectroscopic mosaic contains 11094 spectra covering an
area of $\sim$34 arcmin$^2$ (see papers I and II for details),
which is the  largest area ever covered by an IFU mosaicking.
The V300 grating was used for all the observations, covering the wavelength range $\sim$3700-7100\AA, 
with a spectral resolution of FWHM$\sim$8\AA. 
The seeing varied between $\sim$1 and $\sim$1.8 arcsec, and the median seeing 
was 1.4 arcsec. Different spectrophotometric stars were observed during the observing
runs with, at least, two stars observed each night. The estimated spectrophotometric
accuracy is $\sim$0.2 mag.
Figure~\ref{mosaico_n628} shows the mosaic pattern covered by the
observations. 
The data reduction is described in Rosales-Ortega et al. (2010)\nocite{Rosales-Ortegaetal2010} and
will not be repeated here.  The main properties of the galaxy are
shown  in Table~\ref{tab:n628}.

\begin{table*}
\begin{tabular}{lcccccccc}
\hline\hline
  \begingroup\centering Object\endgroup  
& \begingroup\centering Type  \endgroup
& \begingroup\centering Distance \endgroup
& \begingroup\centering projected size \endgroup
& \begingroup\centering M$_B$  \endgroup 
& \begingroup\centering i   \endgroup 
& \begingroup\centering PA  \endgroup  
& \begingroup\centering r$_{25}$ \endgroup     
& \begingroup\centering r$_d$ \endgroup\\
        &          & \begingroup\centering (Mpc)\endgroup   
                   & \begingroup\centering (arcmin) \endgroup       
                   & \begingroup\centering mag   \endgroup
                   & \begingroup\centering degrees \endgroup
                   & \begingroup\centering degrees \endgroup
                   & \begingroup\centering (arcmin) \endgroup 
                   & \begingroup\centering (arcsec) \endgroup    \\ 
\hline
NGC628  & SA(s)c   &  9.3         &10.5$\times$9.5  & $-19.9$ &  6     & 25     &  5.23        &   71.8 (1)  \\  
\hline
\end{tabular}
\caption{Main properties of NGC~628. \label{tab:n628}}
\end{table*}

\section{Analysis}
\label{sec:analysis}
\subsection{Emission line removal}

We used a clean version of the integral field spectroscopy (IFS hereafter) mosaic of NGC 628 
obtained by: 1) applying a flux threshold
 cut choosing only those fibers with an average flux along the whole spectral range greater 
 than 10$^{16}$ ergs$^{-1}$ cm$^{-2}$ \AA$^{-1}$
 (in order to get rid of spectra for which no information could be derived); 2) removing 
 bad fibers (due to cosmic rays and CCD cosmetic defects) and 
foreground stars (10 within the observed FOV of NGC~628). 
In total, 63\% of clean fibers were preserved with respect to the original mosaic.

The first step to analyse the stellar population properties is to decouple the emission from the underlying stellar population.
 One of the advantages of IFS over the long slit observations is that it allows this decoupling in a spatially-resolved 
basis. The 
underlying stellar continuum was thus decoupled from the emission lines following the procedure outlined in papers I and II. 
First, for each fiber, the stellar population was fit by a linear combination of synthetic  templates of three ages 
(0.09, 1.00 and 17.78 Gyr) and two metallicities (Z $\sim$ 0.0004 and 0.03) from the MILES library 
(Vazdekis et al. 2010)\nocite{Vazdekisetal.2010} using FIT3D (S\'anchez et al. 2012)\nocite{Sanchezetal.2012}. These 
templates are too simplistic to describe in detail the stellar populations in 
the dataset, and were merely used to get a first-order model of the stellar continuum to be subtracted from each spectrum, 
obtaining a pure-emission one. 

Based on this residual spectrum, we obtained a model for the most prominent emission lines in the considered wavelength range, 
including: H$\alpha$, H$\beta$, H$\gamma$, [O II] $\lambda$3727, [O III] $\lambda\lambda$4959,5007, 
[N II] $\lambda\lambda$6548,6583 and [SII] $\lambda\lambda$6717,6731, and also for the most prominent sky residuals 
present in the spectrum, 
i.e. [O I] $\lambda$5577, and [Na I] $\lambda$5893 lines. The modeled emission lines were subtracted from the observed 
spectrum, obtaining an 
emission-free one. Finally, the data was spatially resampled to a data cube with a regular grid of 2 arcsec/spaxel, 
adopting a flux-conserving 
interpolation method, leading to a pure-continuum 3D cube.

\begin{figure}
\centering
\resizebox{0.45\textwidth}{!}{\includegraphics[angle=-90]{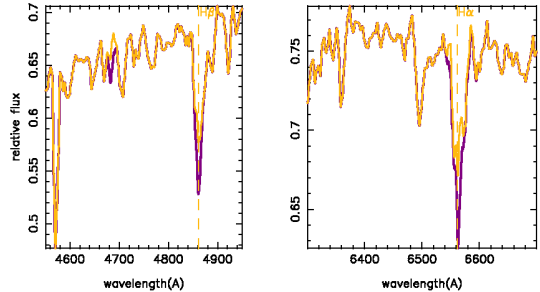}}
\caption{Typical residual emission in the spectra after removing the
  HII regions from the data cube. The original spectrum is labeled
  with orange while the purple show the spectra after subtraction of
  the nebular emission. \label{emission_example}}
\end{figure}

However, the aim of this removal in papers I and II was to extract the strong emission lines from HII regions, not the diffuse 
component and, with this aim, a threshold in the emission line flux was imposed. In the reanalysis of the 
data presented here, we have also  removed the fainter diffuse emission remaining in some of the 
spectra.
To do  this, we use the software {\tt GANDALF}
(Sarzi et al. 2005)\nocite{Sarzietal.2005}\footnote{The use of a different code simply does not respond to any reason in particular,
and this second step could have been perform also with FIT3D.}. {\tt GANDALF} fits, simultaneously, the absorption and 
emission lines, treating the
latter as additional Gaussians. In this respect, the procedure is very similar to the one carried out with FIT3D.
In a first step, emission lines are masked and the absorption line
spectrum is fit with the penalized pixel-fitting 
{\tt pPXF} (Cappellari \& Emsellem 2004\nocite{Cappellari&Emsellem2004}), using as
templates the new stellar population models of Vazdekis et al. (2010)\nocite{Vazdekisetal.2010} based on the MILES library
(S\'anchez-Bl\'azquez et al. 2006\nocite{Sanchez-Blazquezetal2006d}; Cenarro et al. 2007\nocite{Cenarroetal.2007})
\footnote{http://miles.iac.es}. 
In this step, radial
velocities and velocity dispersions ($\sigma$, hereafter) for the stellar component are derived. The best
values of velocity and $\sigma$ and the best template mix are then used as initial values for the
derivation of emission lines using {\tt GANDALF}. Emission line equivalent-widths, radial velocities and
$\sigma$ for the gaseous component are derived in this second step. The fit allows for a low-order Legendre
polynomial in order to account for small differences in the continuum shape between the pixel spectra and
the templates. The best fitting template mix is determined by a $\chi^2$ minimization in pixel space.
Emission line spectra at each bin were subtracted from the observed spectra for the subsequent analysis.
Figure~\ref{emission_example} shows an example of the residual emission
in two of the spectra before and after removing the nebular component.

\subsection{Star formation histories}
\label{sec:steckmap}
The most popular approach to derive stellar population properties from
integrated spectra has been to compare absorption lines --usually
Lick/IDS indices-- with those computed with single stellar populations (SSP) of a given 
age and metallicity (e.g., Trager et al. 2000\nocite{Trageretal.2000}; 
Thomas et al. 2005\nocite{Thomasetal.2005}; S\'anchez-Bl\'azquez et
al. 2006abc\nocite{Sanchez-Blazquezetal2006a}\nocite{Sanchez-Blazquezetal2006b}
\nocite{Sanchez-Blazquezetal2006c}; Kuntschner et al.\ 2006\nocite{Kuntschneretal2006}, among many others). Although
this may not be a bad approximation for luminous, early-type galaxies,
studies of resolved stellar  population in nearby discs have shown 
that their star formation history is better represented by
exponential shapes, and that 
enhanced star formation at later times is 
not uncommon (e.g., Kennicutt, Tamblyn \& Congdon 1994\nocite{KTC94}; Huang et al. 2013\nocite{Huangetal2013}). 
Therefore, the SSP approach is not adequate for the present study
\footnote{However, for the sake of comparison with other studies we present the Lick/IDS line-strength indices maps in 
Appendix~\ref{appendix_lickmap}.}.  Furthermore, 
we want to study the possibility of deriving, not only mean ages and metallicities, but also the evolution of the 
metallicity gradient with time, for which we need to derive the whole star formation history and the evolution 
of the metallicity with age in all spectra.

In this paper we have chosen to perform a full spectral fitting using the code {\tt STECKMAP}  
(STEllar Content  via Maximum A Posteriori likelihood, Ocvirk et al.\ 2006ab
\nocite{Ocvirketal.2006a}\nocite{Ocvirketal.2006a}\footnote{http://astro.u-strasbg.fr/$\sim$ocvirk/}). 
In {\tt STECKMAP},  the reconstructions of the stellar age distribution and the age-metallicity relation 
are non-parametric, i.e. no specific shape is assumed. The only {\it a priori} conditions that we use are positivity and the requirement
that the solution (the variation of the flux with age and the age metallicity relation) 
is sufficiently smooth. The smoothness parameter can be set by generalized cross-validation 
according to the level of
noise in the data in order to avoid over-interpretation. 
We use here the stellar population models MILES,
spanning an age range from 6.3$\times$10$^7$ to 1.7$\times$10$^{10}$ yr, divided in 30 logarithmic age bins, and a metallicity range
$[$Z/H$]$=$[+0.2$,$-1.3]$.

The function to minimize (the objective function) is defined as:
\begin{equation}
Q\mu=\chi^2 (s(x,Z,g))+P_{\mu}(x,Z,g),
\end{equation}
which is a penalised $\chi^2$, where $s$ is the model spectrum, $x$ represent the flux  distribution, $Z$ the metallicity 
distribution and $g$ the broadening function. The penalization $P_{\mu}$ can be written as: $P_{\mu}(x,Z,g)=
\mu_x P(x)+\mu_z P(Z)+\mu_v P(g)$, where the function $P$ gives high values for solutions with strong oscillations  (ie., 
when the flux or the metallicity changes rapidly with time) and small values for smoothly varying solutions. Adding 
the penalization $P$ to the objective function is exactly equivalent to injecting {\it a priori} 
probability density to the solution as $f_{\rm prior}(x)=e^{-\mu_x P(x)}$
(see Ocvirk et~al. 2006ab for more details), where $P$ is a quadratic function of the unknown.

Choosing the right values for the smoothing parameters $\mu_{x,z}$ 
is not a trivial problem. In principle, one could choose the values
giving the smaller $\chi^2$ in the fit, but this usually yields a wide range of smoothing parameters, spanning typically 3-4
decades, in which the fit is acceptable. In any case, the exact choice of the smoothing parameter, although  
affecting the detailed shape 
of the star formation history and the
age-metallicity relation, does not impact the overall interpretation of the fit, and we have checked, via Monte Carlo
simulations, that the global shape of the solution and, in particular, the average values, such as the luminosity-weighted and
mass-weighted age and metallicity for the old and young components  are stable for this range of smoothing parameters. The final chosen values were $\mu_{x,z}=1$.

Although the code allows to
fit the kinematic properties of the galaxy simultaneously with its stellar content, 
for this work we have not fit the kinematics, and we use, instead,
the values obtained with ${\tt ppxf}$ (Cappellari \& Emsellem 2004\nocite{Cappellari&Emsellem2004})
during the emission line removal.
The reasons are explained in more detail in Appendix~B of S\'anchez-B\'azquez et al.~2011\nocite{Sanchez-Blazquezetal.2011}. Basically, the existing degeneracy
between the metallicity and the velocity dispersion (Koleva et al. 2008)\nocite{Kolevaetal2008} biases the mean-weighted 
metallicity if both parameters are
fitted at the same time. We have found that this degeneracy affects, mostly, values of the metallicity, in the
sense that the derived mass-weighted metallicity is higher if the kinematic is not fixed\footnote{Note that this systematic 
effect depend on the specific dataset. In fact, the opposite trend was found in S\'anchez-Bl\'azquez et al. 2011}. 
Furthermore, in this case, 
the program fails to recover  properly the age-metallicity relation (see S\'anchez-Bl\'azquez et al.~2011). However, the
biases disapear when using spectra of high signal-to-noise ($>70$)
Appendix~\ref{appendix2} shows the comparison of the derived mean stellar population parameters with and without fixing the 
kinematics. 

\begin{figure*}
\centering
\resizebox{0.7\textwidth}{!}{\includegraphics[angle=-90]{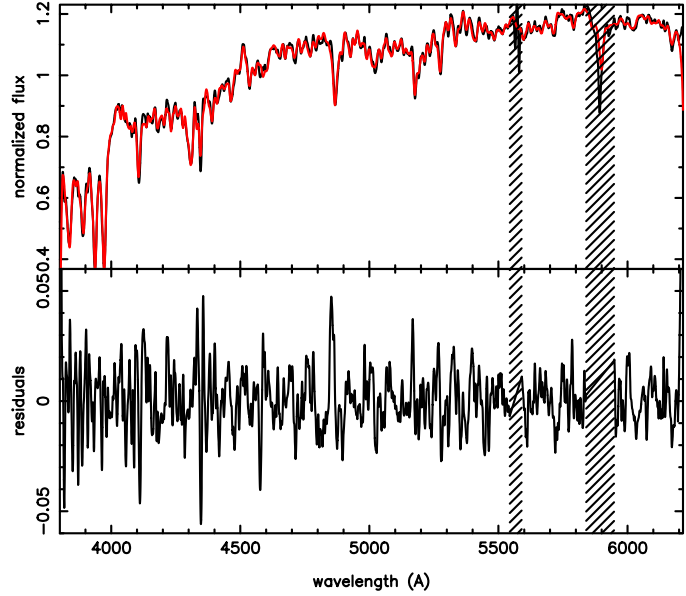}}
\caption{Example of the fit (red line) obtained with {\tt STECKMAP} for the integrated spectrum of the galaxy (black line).
The bottom panel shows the residuals from the fit (observed-fitted). Note the scale of this panel. 
Hashed regions indicate the regions that were  masked during the fit.\label{steckmap_fits}}
\end{figure*}

\begin{figure*}
\centering
\resizebox{0.32\textwidth}{!}{\includegraphics[angle=-90]{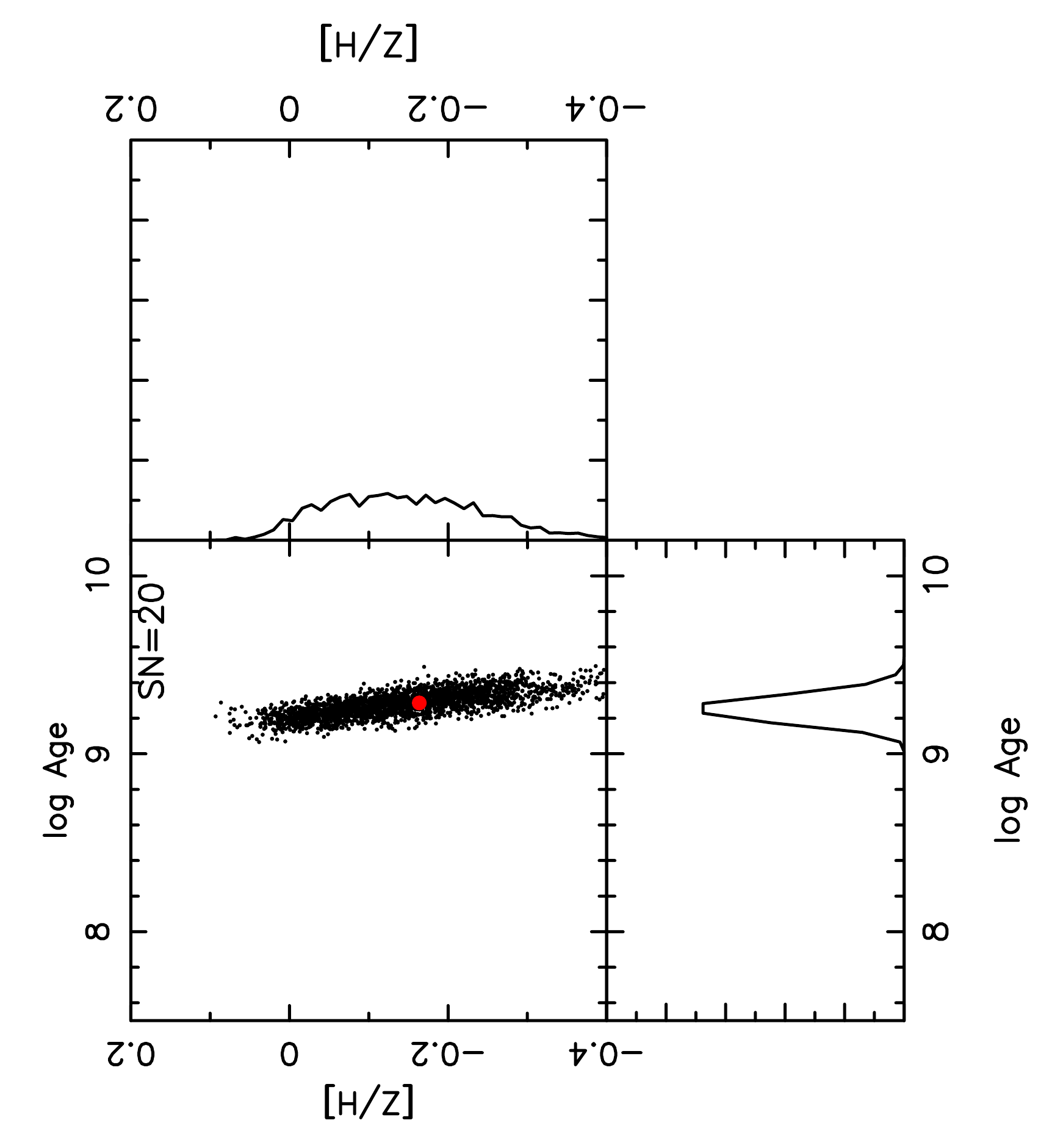}}
\resizebox{0.32\textwidth}{!}{\includegraphics[angle=-90]{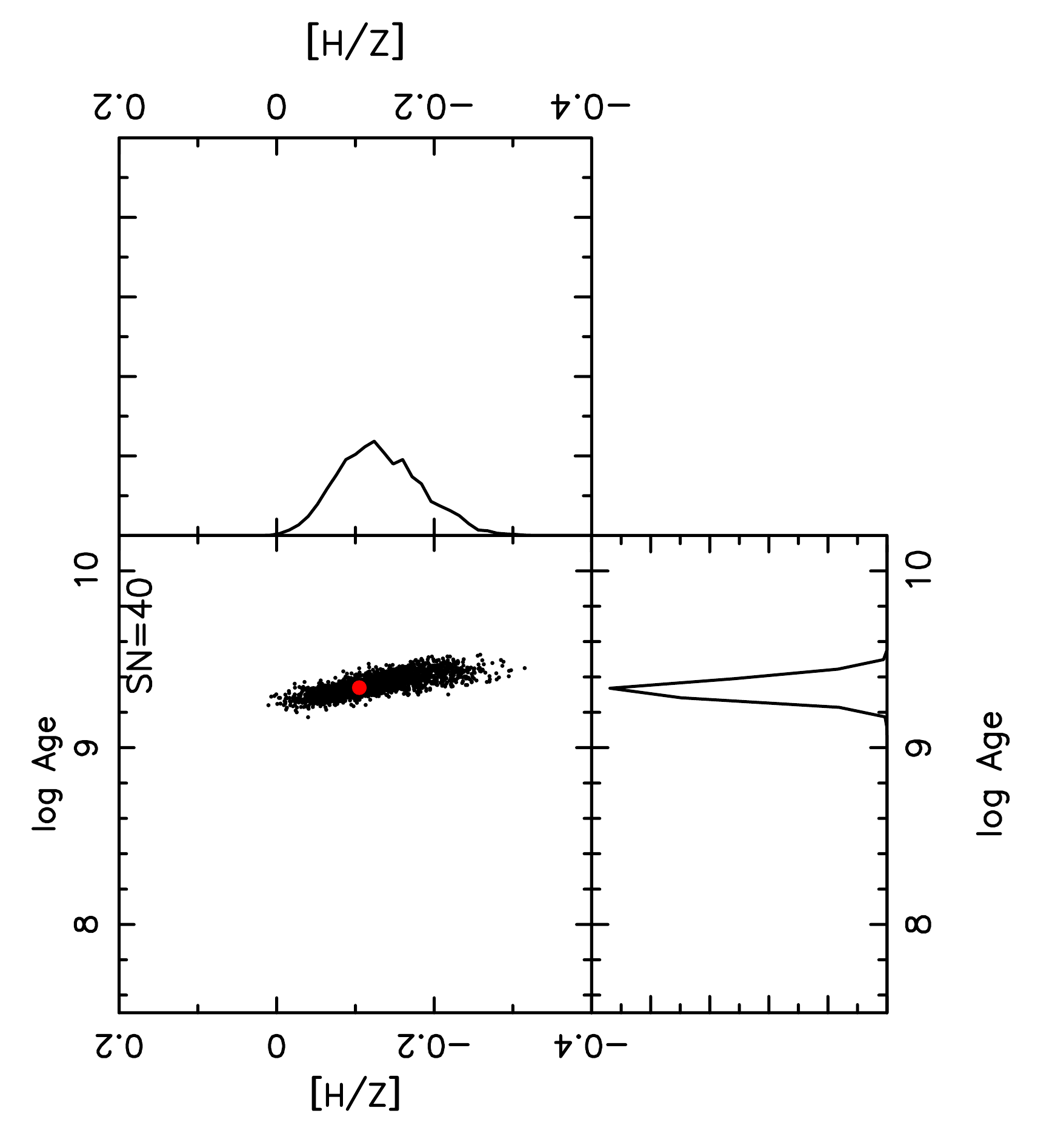}}
\resizebox{0.32\textwidth}{!}{\includegraphics[angle=-90]{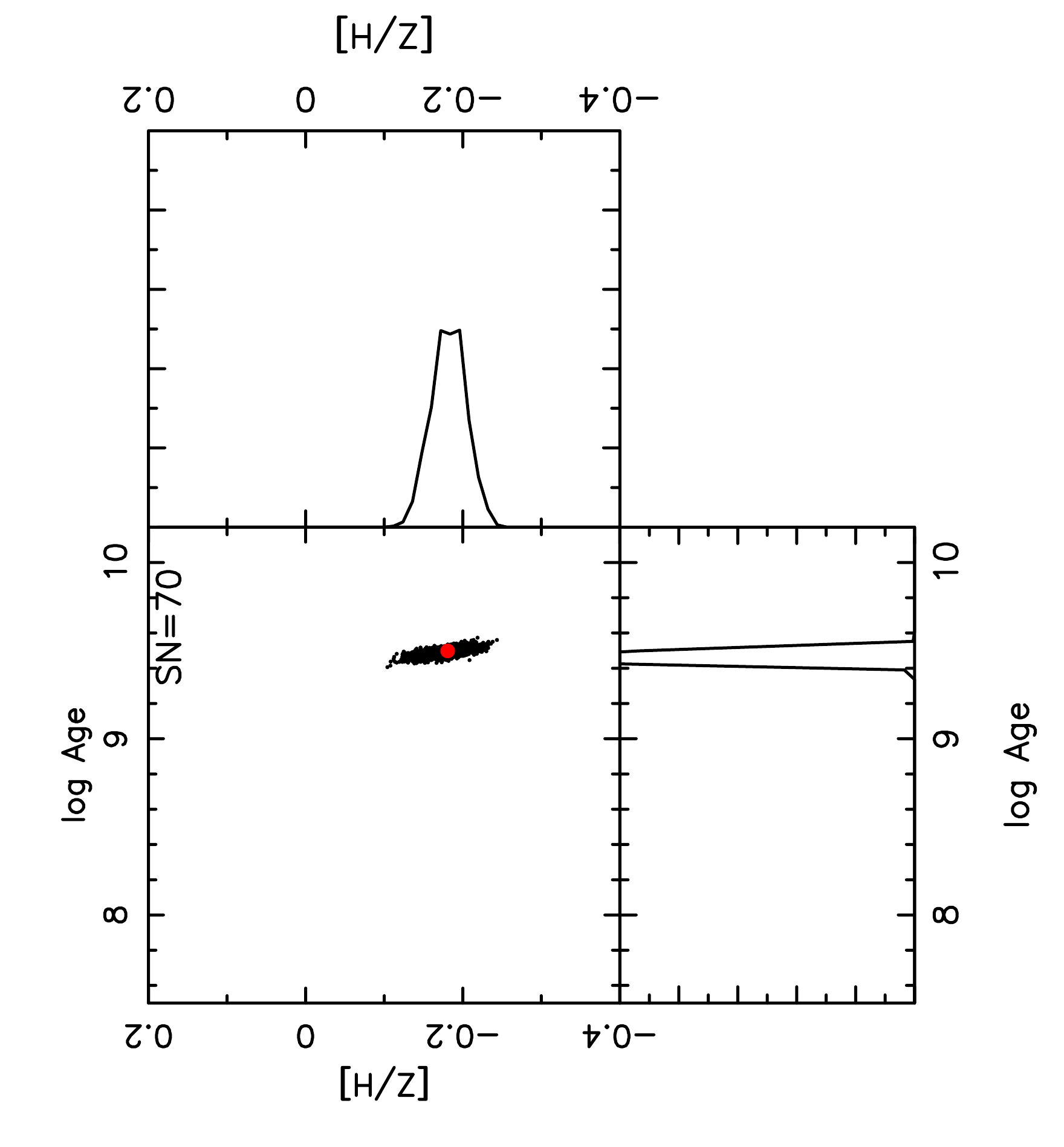}}
\caption{Luminosity-weighted age-metallicity relation obtained by performing 2500 Montecarlo simulations in which each pixel is 
perturbed with the associated noise error, following an Gaussian distribution. The histograms show the distribution 
of the mean age and metallicity obtained in each simulation. From left to right, test performed for spectra
of signal-to-noise 20, 40 and 70.\label{chi_map}}
\end{figure*}

In order to deal with possible flux calibration errors, we multiply the model by a smooth non-parametric transmission curve,
representing the instrumental response multiplied by the interstellar extinction. This curve has 30 nodes spread uniformly along
the wavelength range, and the transmission curve is obtained by spline-interpolating between the nodes. The latter are treated as
additional parameters and adjusted during the minimization procedure. This continuum matching technique is similar in essence to the
multiplicative polynomial used by {\tt NBurst} (Chilingarian et al.\ 2007)\nocite{Chilingarianetal2007}; {\tt Gandalf} 
(Cappellari \& Emsellem 2004)\nocite{Cappellari&Emsellem2004}  or {\tt ULySS} 
(Koleva et al. 2009)\nocite{Koleva2009}.  This decision 
of ignoring  the continuum shape in the fit  was taken to avoid biases in the derivation 
of star formation histories due to possible flux calibration errors, 
to a non-perfect extinction correction  or to any other cause affecting the shape of
the continuum.  Appendix~\ref{appendix_cont} shows the 
comparison of the  mean stellar population parameters obtained without using the shape 
of the continuum and 
those obtained when we allow the continuum 
to be fitted. 
As can be seen, fitting or not the continuum have a very strong influence in the derived parameters, specially in the mean metallicities.
In general, the metallicities obtained when the continuum is rectified are higher than those obtained when we fit the continuum. 
This is not necessarily 
a rule as it will depend on the particular shape of the continuum, affected by the particular data flux calibration, extinction correction, etc.
However, this is only true for lower signal-to-noise data.  For  spectra with signal-to-noise (per anstrong) higher than 30 the values obtained 
are independent of our choice of using or not the continuum for our analysis. 
We fit the spectral range 3800-6215\AA, masking the region between 5545 and 5590~\AA, affected by sky-removal residuals.
Figure~\ref{steckmap_fits} shows an  example of the fit obtained with {\tt STECKMAP} for the integrated spectrum of the galaxy.
\begin{table*}
\begin{tabular}{lllllllll}
\hline
S/N & MW (log Age)& RMS    & MW ([Z/H]) & RMS   & LW (log Age) & RMS    & LW ([Z/H]) & RMS  \\  
\hline\hline
70  & 9.976       &  0.015 &  $-0.285$    & 0.035 &  9.491       & 0.021  & $-0.176$     & 0.022\\
40  & 9.816       &  0.079 &  $-0.204$    & 0.062 &  9.359       & 0.054  & $-0.125$     & 0.052\\
20  & 9.706       &  0.081 &  $-0.263$    & 0.125 &  9.281       & 0.070  & $-0.143$     & 0.099\\
\hline
\end{tabular}
\caption{Luminosity-weighted (LW) and mass-weighted (MW) mean values and RMS resulting of 2500 simulations where
each pixel was perturbed with Gaussian noise expected for a spectrum with a signal-to-noise per angstrom of 70, 40 and 20. \label{table:simus}}
\end{table*}
One of the most important problems affecting stellar population studies is the well known age-metallicity 
degeneracy, i.e.., the colours and spectral characteristics of a given population can be mimicked with another population 
younger but more metal rich or older but more metal poor. To explore upon what extent we are affected
by this degeneracy, we use the integrated spectrum of NGC~628 and added noise to simulate spectra of 
signal-to-noise 20, 40 and 70 (70 is the S/N of the original spectrum). We performed 2500 Montecarlo simulations 
in which each pixel was perturbed with the associated noise error, following a Gaussian distribution. For each 
simulation a new value of the age and metallicity was derived. The results are shown in Fig.~\ref{chi_map}. 
The mean values obtained and the root-mean square dispersion (RMS) are shown in Table~\ref{table:simus}.
As can be seen, as the noise increase, also the uncertainty in the parameters and, therefore, the age-metallicity 
degeneracy. Furthermore, there seem to be a small systematic effect in the sense that as the noise increase, the 
mean derived age decrease. 
We decided to bin our data with a minimum signal-to-noise of 40 (see Sec.~\ref{sec_maps}) as a compromise
between obtaining good quality spectra and resolve spatially the different components of the galaxy.
The typical dispersion  in the mean luminosity-weighted age and metallicity derived in a spectrum with 
this signal-to-noise is  0.05, and slightly larger for the mass-weighted means 
(0.08 and 0.06 for the age and metallicity values respectively).

\begin{figure*}
\centering
\resizebox{0.7\textwidth}{!}{\includegraphics[angle=-90]{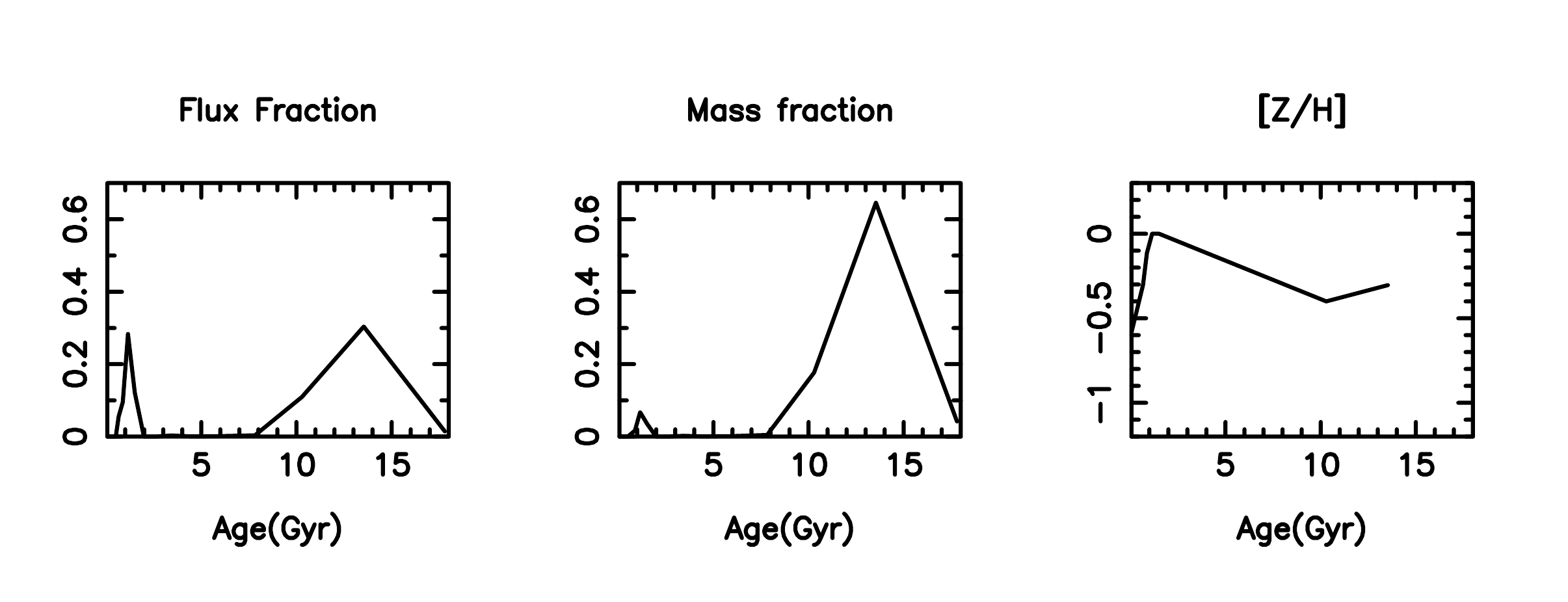}}
\caption{Mass and flux-fractions derived with {\tt STECKMAP} for the integrated spectrum of NGC~628. \label{fig_sfr}}
\end{figure*}

Figure~\ref{fig_sfr} shows the typical {\tt STECKMAP} outputs for the integrated spectra of NGC~628. The flux and mass fraction 
and the derived age metallicity relation shows that the majority of the stars in this galaxy formed $~$13 Gyr ago and less than 
10\% of stars did it $\sim$ 1 Gyr ago. Note, that the maximum age of the models used here is 17.78~Gyr, ie., exceeding the 
age of the Universe in the adopted cosmology. It goes beyond the scope of this paper discussing the reasons for these discrepancies 
and we just caution the reader that the  absolute values of the old populations have to be taken as approximate values. 
In any case, this does not affect any of our conclusions as we will only discuss relative stellar population variations along 
the galactic radius and relative trends between very old and young populations.

\section{Results}
\label{sec_results}
\subsection{Age and metallicity maps}
\label{sec_maps}

\begin{figure}
\centering
\resizebox{0.4\textwidth}{!}{\includegraphics[angle=0]{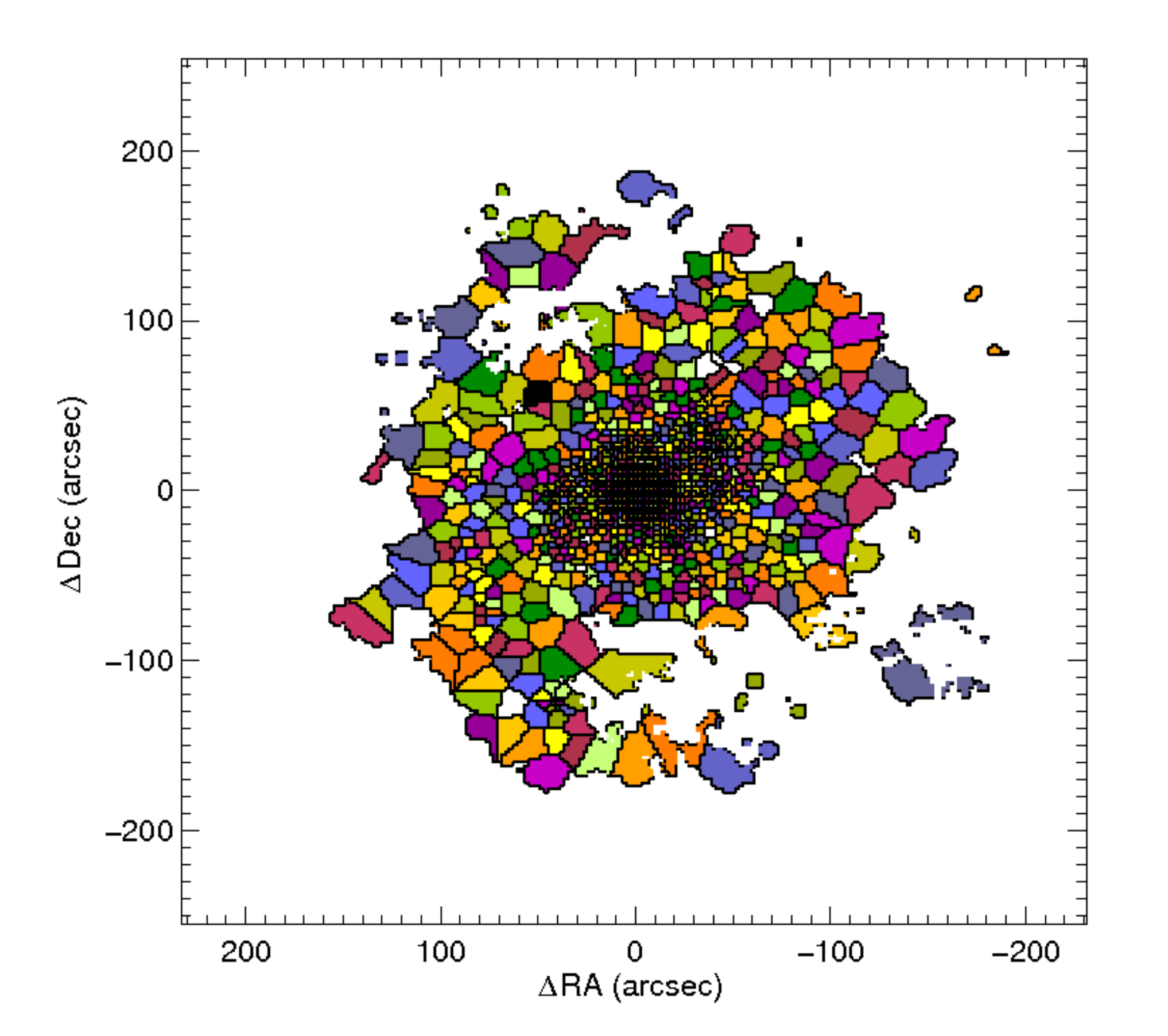}}
\caption{Different regions of the Voronoi-tessellation binning scheme performed in our galaxy. \label{voronoi_binning}}
\end{figure}

In order to reach the desired signal-to-noise ratio for our analysis 
-- to ensure an appropriate modeling of the stellar populations -- we 
used the Voronoi tessellation algorithm of 
Cappellari \& Copin (2003)\nocite{CC03} and the PINGSoft (Rosales-Ortega 2011)\nocite{RosalesOrtega2011}
software to bin the data cube in regions with a minimum signal-to-noise 
(S/N) ratio of 40 (see Sec.~\ref{sec:steckmap}) , over a spectral region centered at 5400~\AA~ with 
a 100~\AA~width.
Figure~\ref{voronoi_binning}  shows the different regions binned in our scheme of 
the Voronoi-tessellation algorithm.

We run {\tt STECKMAP} on each of the binned spectra and obtained star formation histories and 
age-metallicity relations for all of them.
Once we have derived star formation histories, we can obtain average values weighted
by the flux or by the mass of the population. While the mean values weighted with 
mass are going to be biased towards the parameters of the majority of the stars, the 
flux-weighed values are biased to the parameters of the youngest component, as young stars are  
much more luminous in the wavelength range we are studying. 

\begin{figure*}
\resizebox{0.45\textwidth}{!}{\includegraphics[angle=-90]{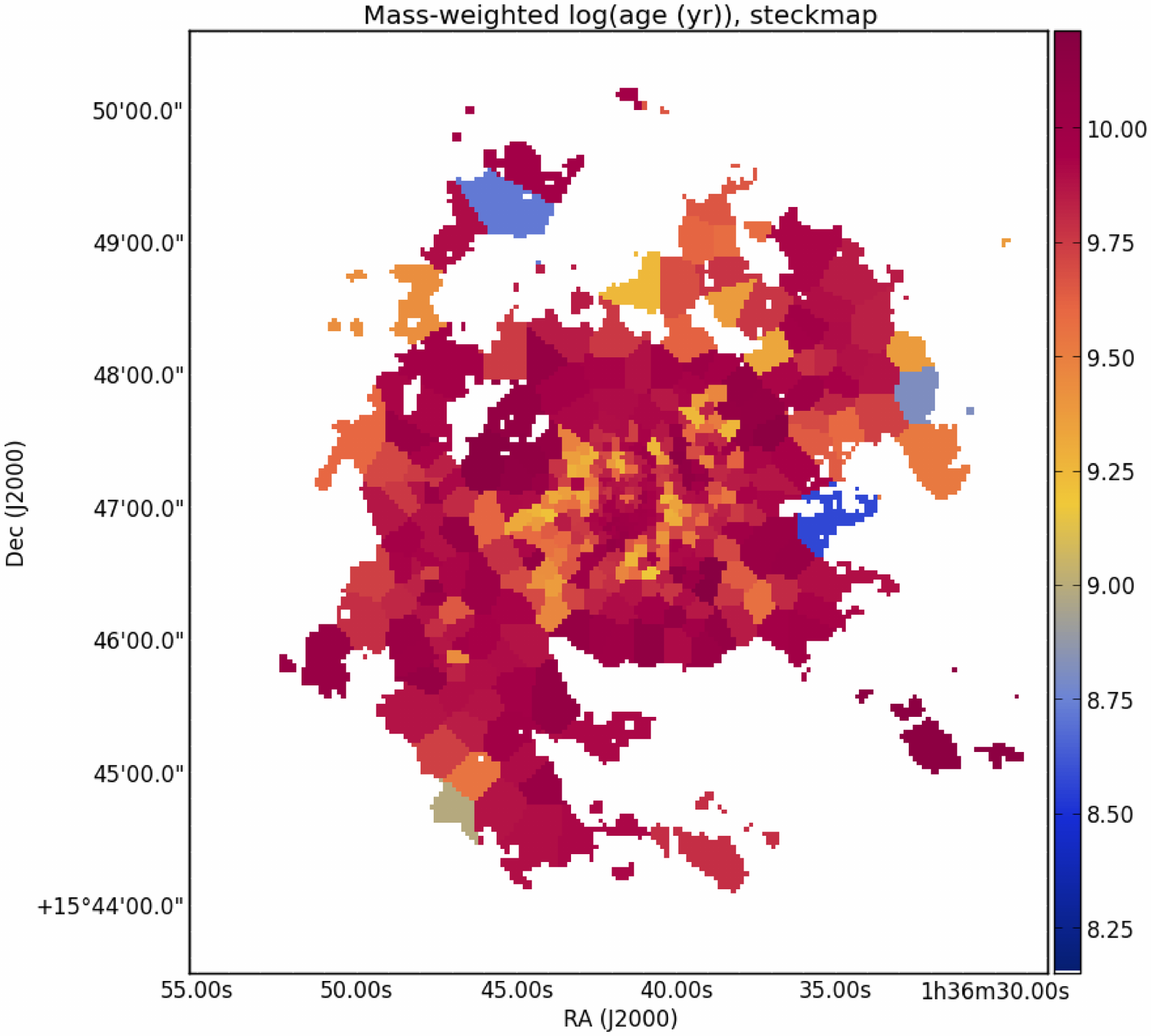}}
\resizebox{0.45\textwidth}{!}{\includegraphics[angle=-90]{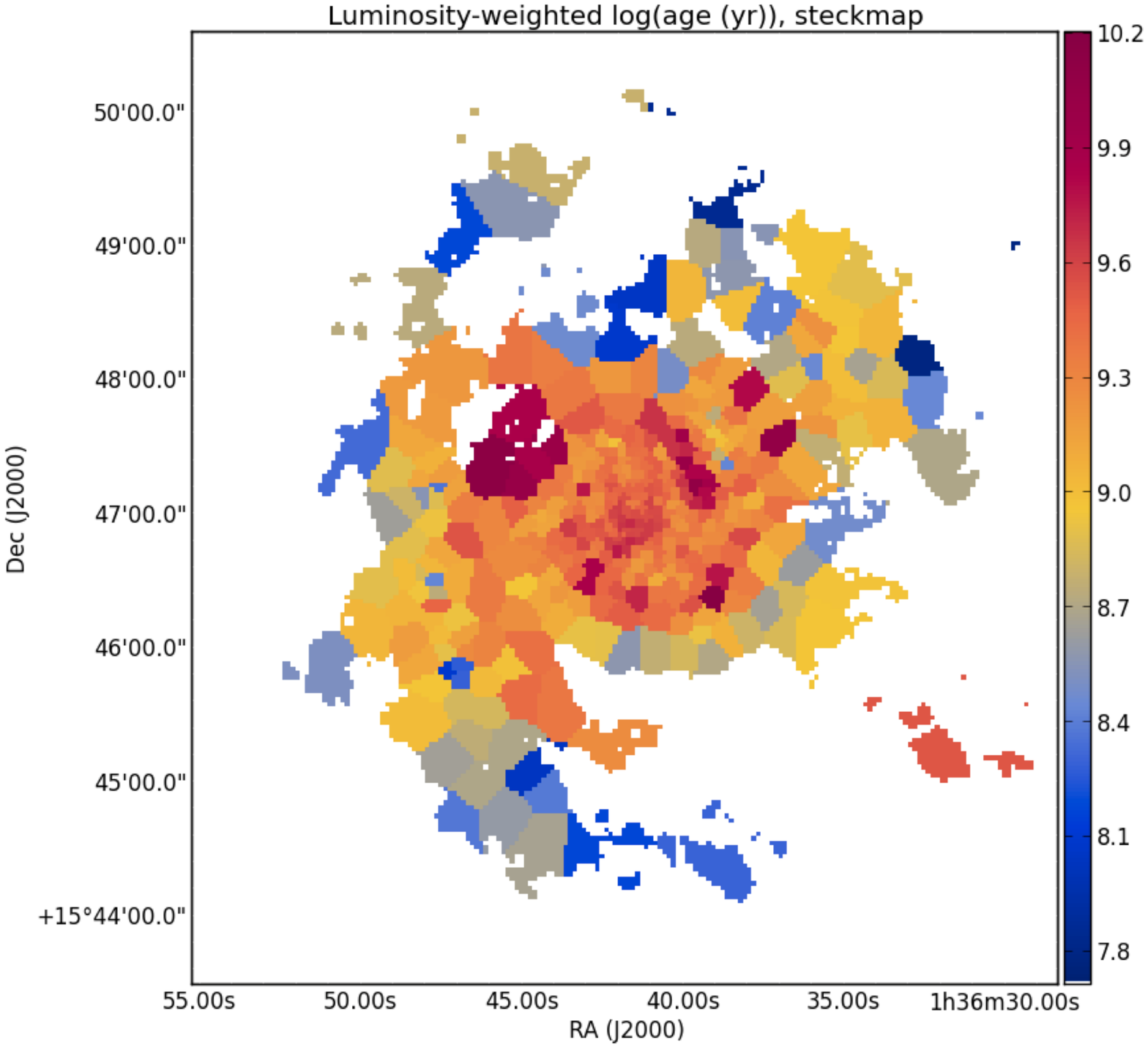}}
\resizebox{0.45\textwidth}{!}{\includegraphics[angle=90]{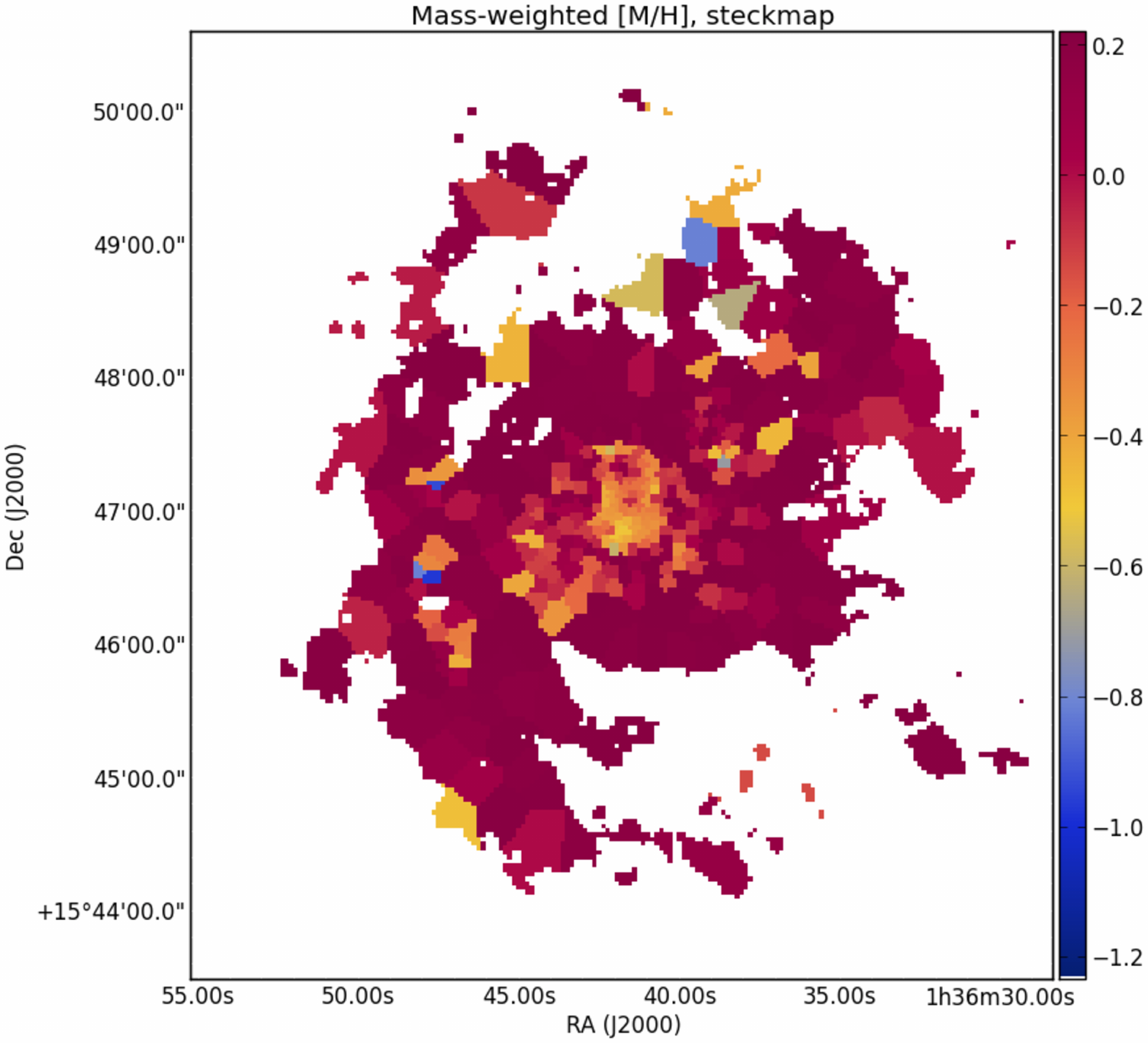}}
\hspace{1cm} \resizebox{0.45\textwidth}{!}{\includegraphics[angle=90]{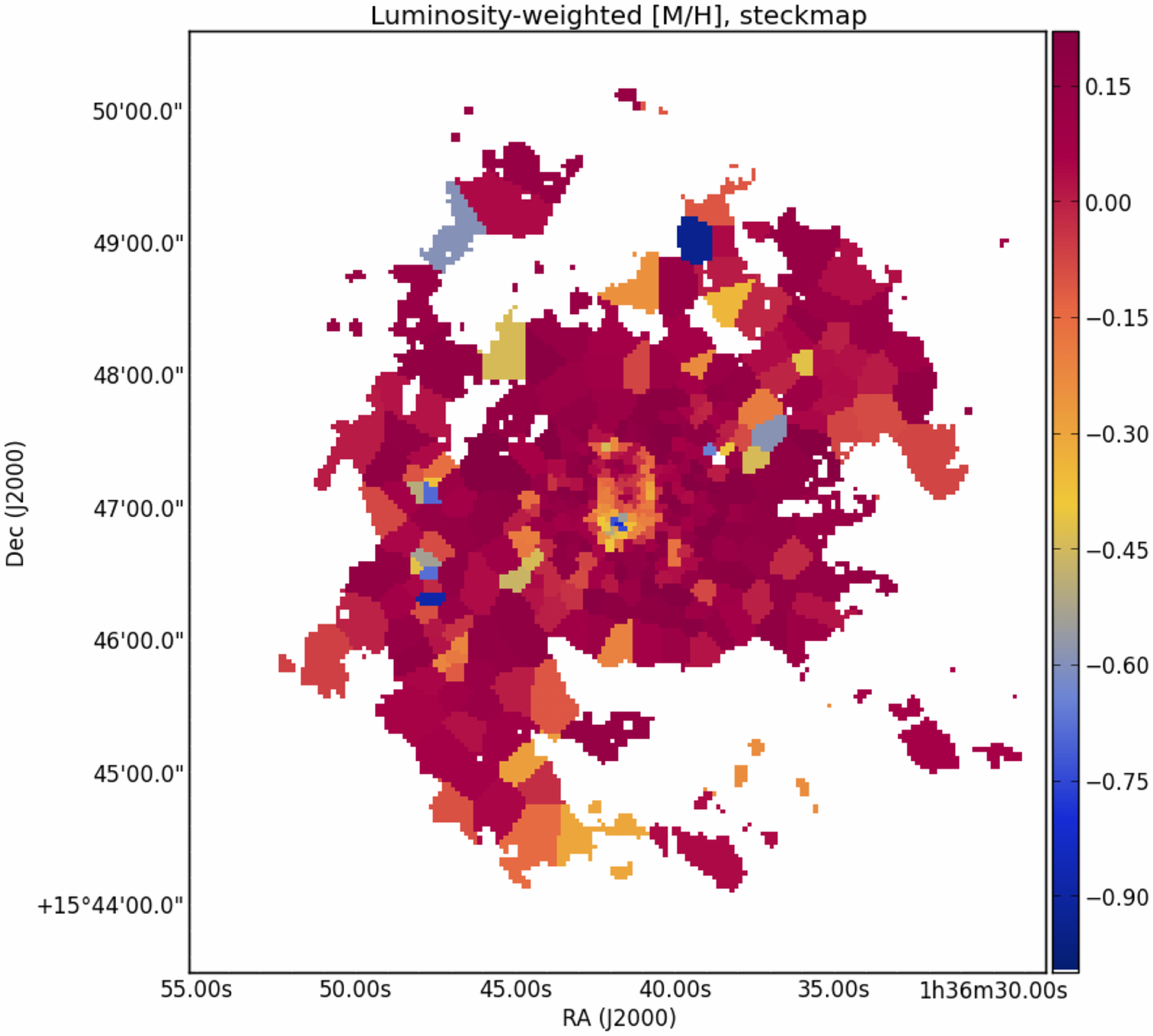}}
\caption{Mean age and metallicity 2-dimensional maps weighted by the mass (MW) and 
by the light (LW) of the stars.\label{fig_maps}}
\end{figure*}

Figure~\ref{fig_maps} shows the luminosity- and mass-weighted maps of age and metallicity
for NGC~628.
Several things can be appreciated in these maps. The mass-weighted mean age map shows that
the majority of the stars in the galaxy are old, even in the most external
regions sampled by the data, while the luminosity weighted age map reveals younger population 
in the external parts. In the mass-weighted age map a ring of younger (with an average age
of $\sim$1.7 Gyr) population is also clearly
visible with yellowish colors, 
coincident with a young ring that has also  been detected in CO (Wakker \& Adler 1995)\nocite{Wakker&Adler1995}.

The mass-weighted metallicity map shows that the ring and the central parts of the galaxy are more metal 
poor than the external parts, which is compatible with the idea that the central ring has been formed as a
consequence of the inflows of gas from the external parts (e.g., Kormendy \& Kennicutt 2004\nocite{KK04}; S\'anchez et al. 2011).
In the next section, we quantify these trends.

\subsection{Stellar population gradients in the disc region}
\label{radialbinning}

To obtain radial trends of the stellar population parameters, we 
average the individual spectra in  concentric annuli, which
allows to  obtain the required signal-to-noise ratio for our analysis up to 
$\sim$ 120 arcsec ($\sim$0.4$\rho_{25}$ or $\sim$5.4 kpc) in linear
projected galactocentric radii. We  obtained azimuthally averaged spectra by co-adding all the spectra
within successive rings of 4 arcsec. 
These azimuthally averaged
spectra were then analysed using the fitting procedure described above.
Following Zou et al. (2011) we will consider the disc the region R$>$30~arcsec, while 
the bulge that of R$<$12.6~arcsec.

\begin{figure*}
\centering
\resizebox{0.3\textwidth}{!}{\includegraphics[angle=-90]{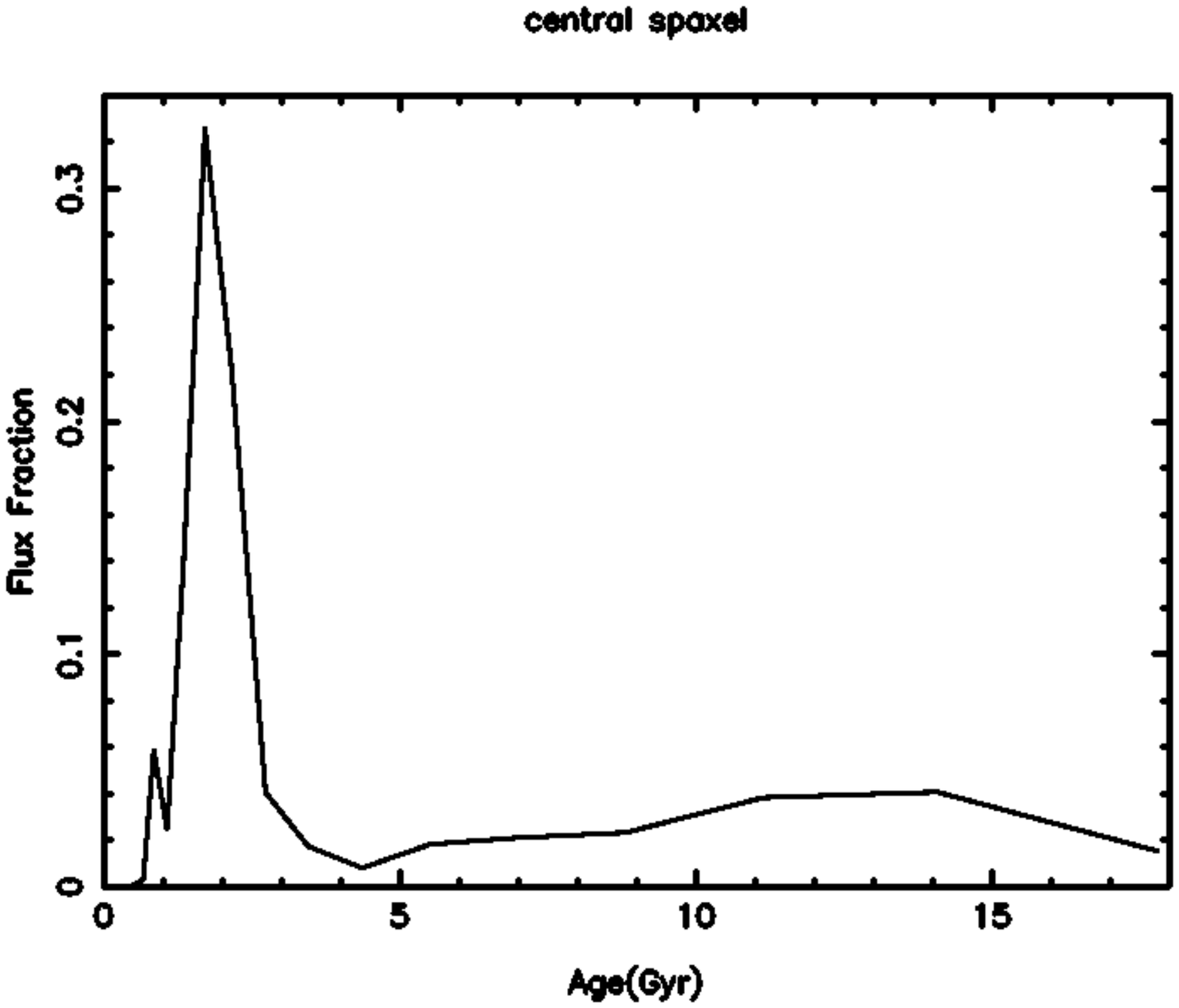}}
\resizebox{0.3\textwidth}{!}{\includegraphics[angle=-90]{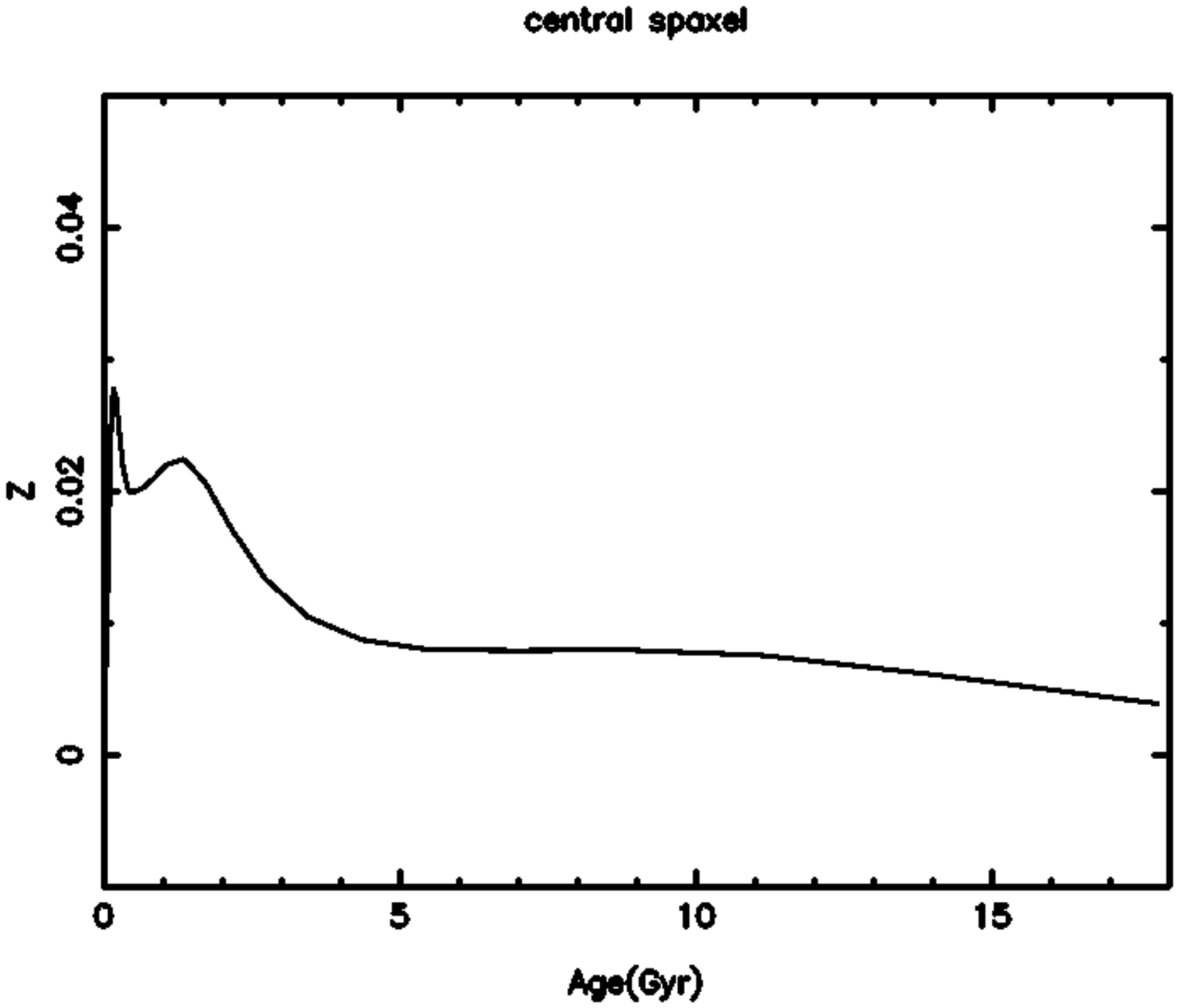}}\\

\resizebox{0.3\textwidth}{!}{\includegraphics[angle=-90]{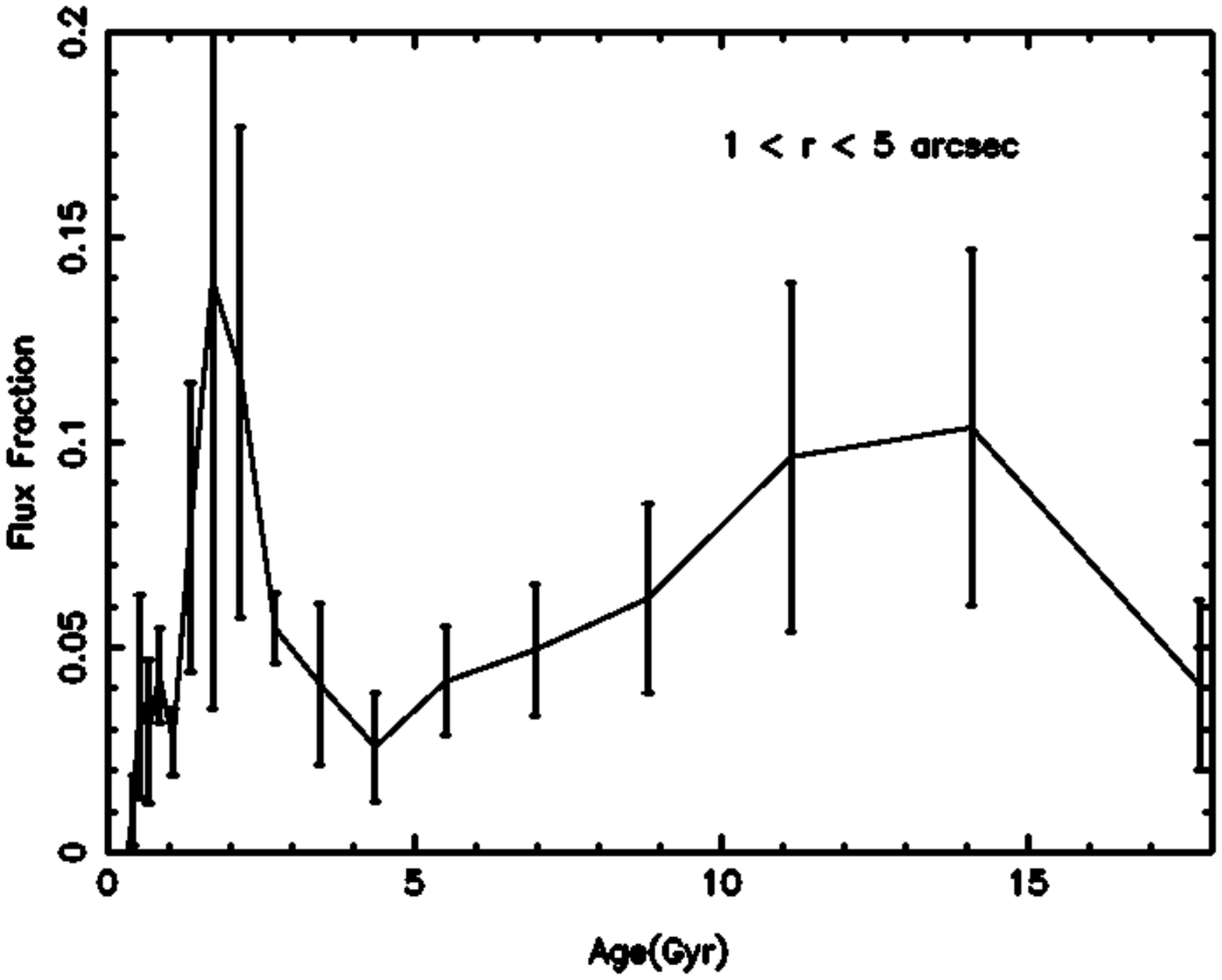}}
\resizebox{0.3\textwidth}{!}{\includegraphics[angle=-90]{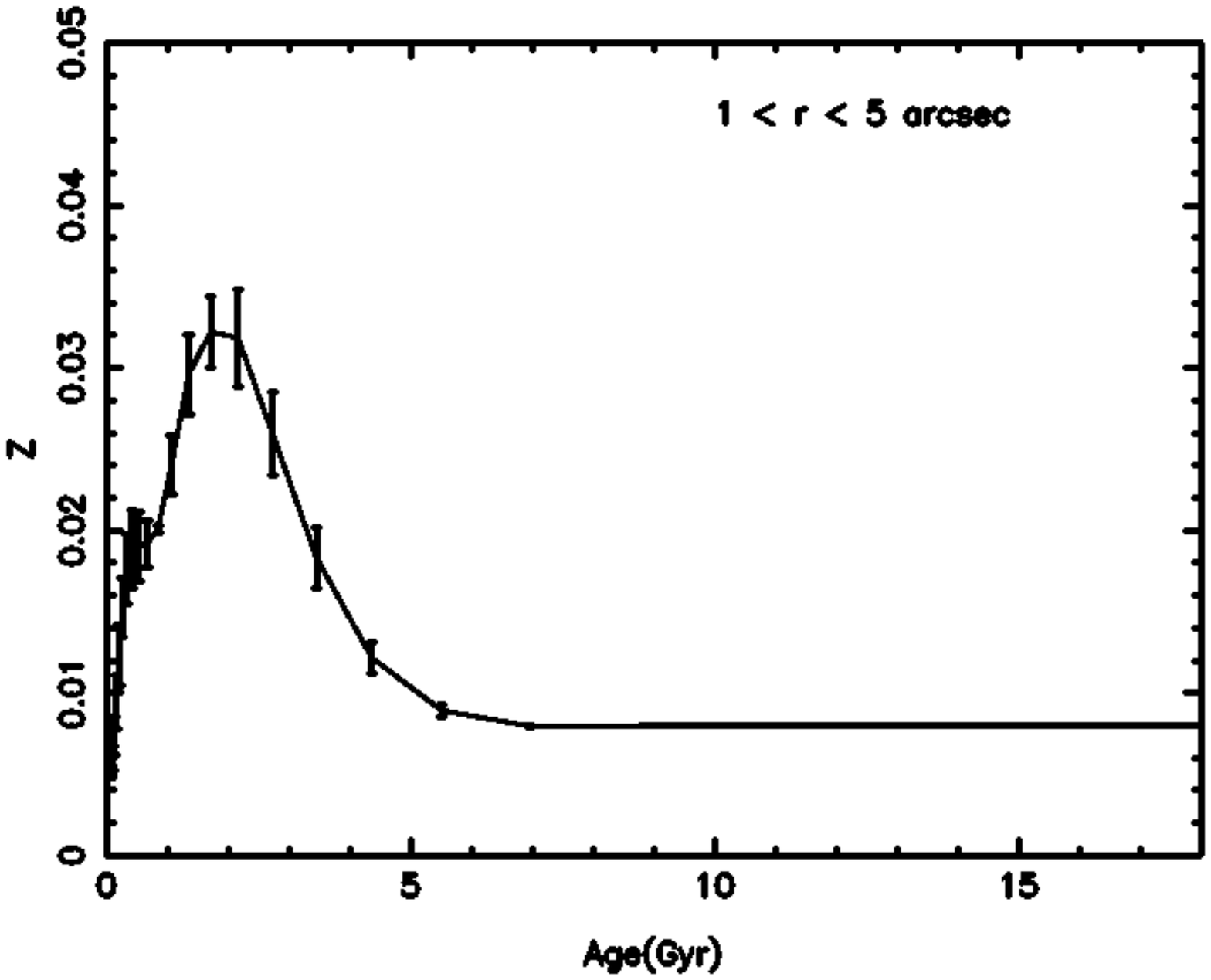}}\\

\resizebox{0.3\textwidth}{!}{\includegraphics[angle=-90]{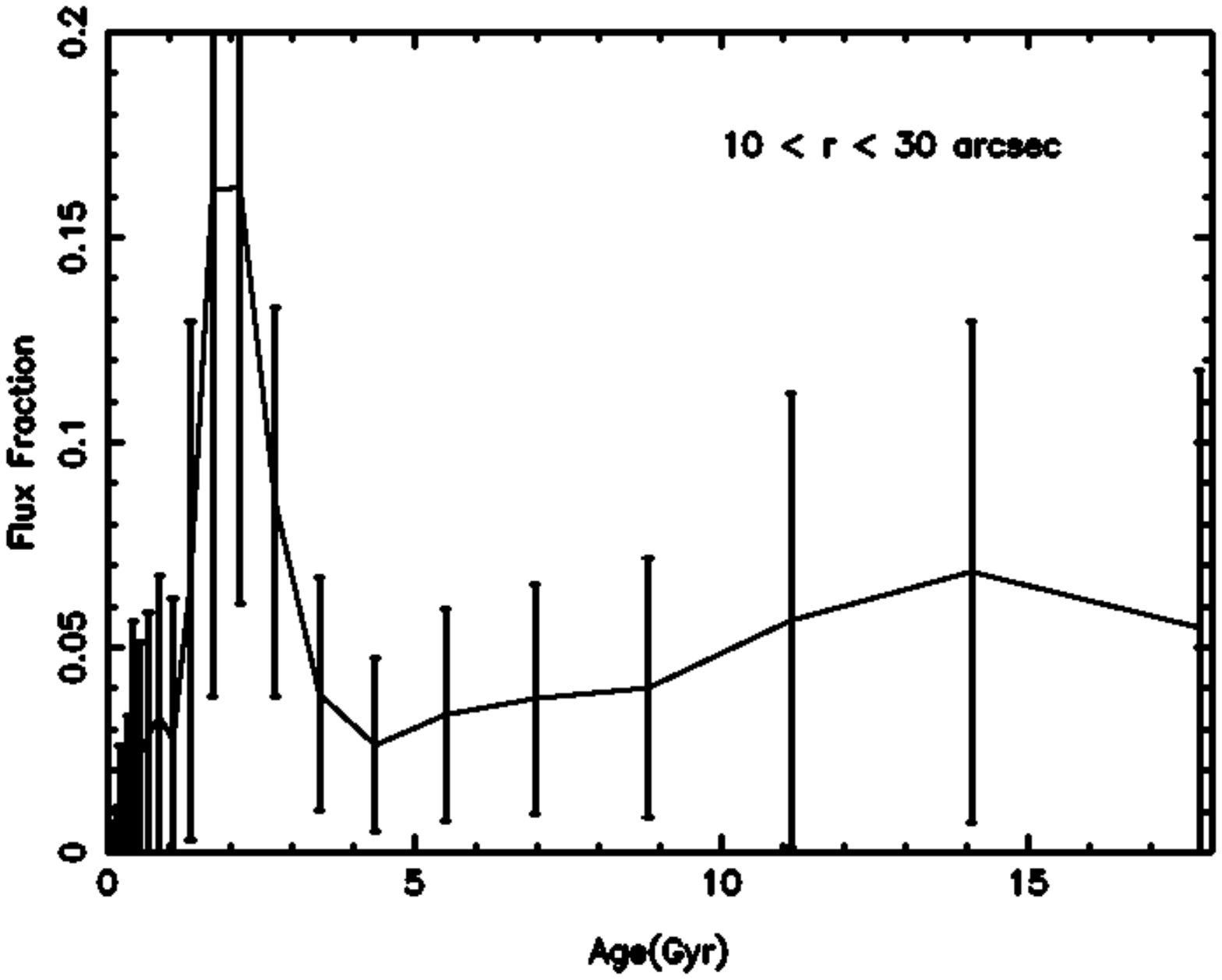}}
\resizebox{0.3\textwidth}{!}{\includegraphics[angle=-90]{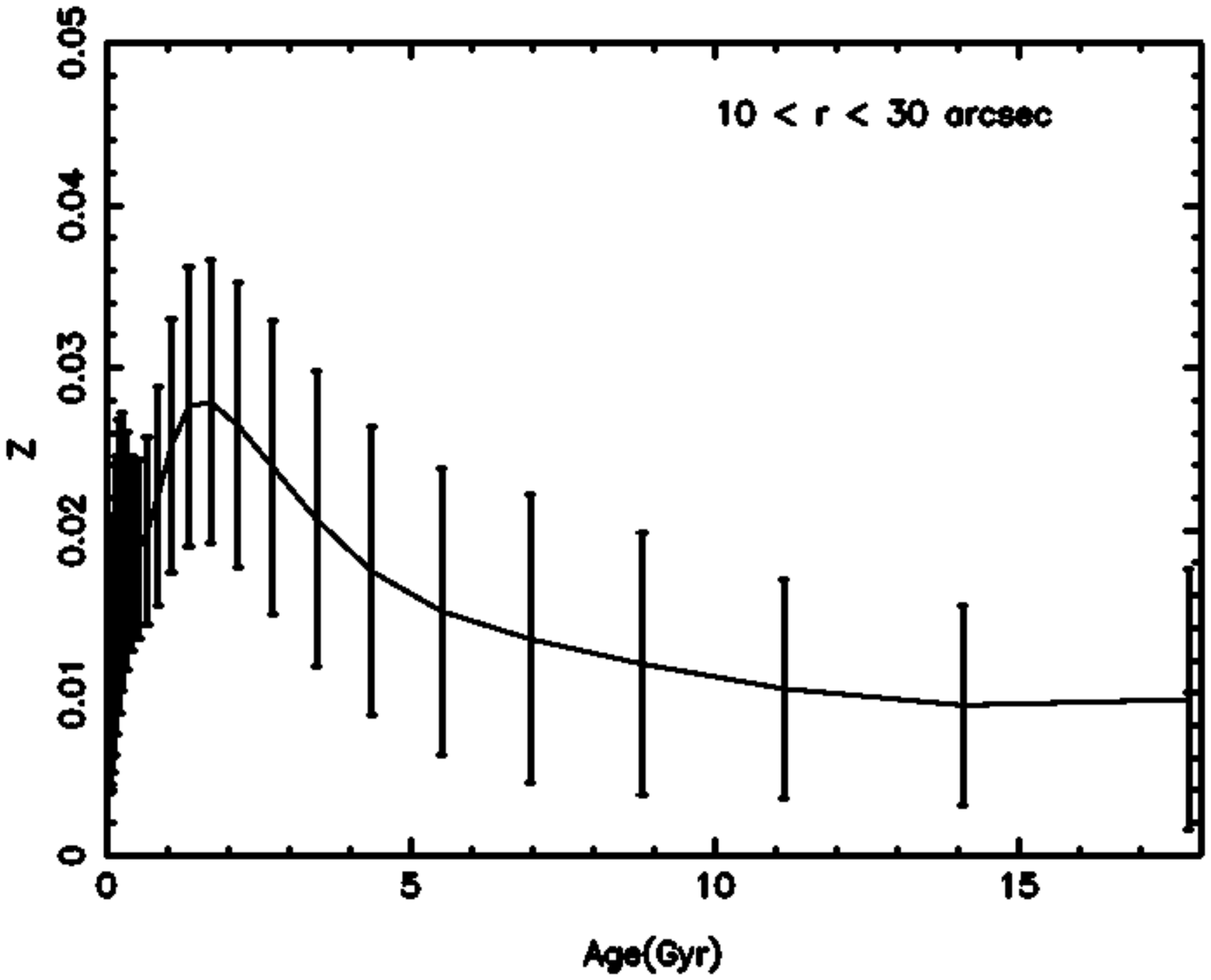}}\\

\resizebox{0.3\textwidth}{!}{\includegraphics[angle=-90]{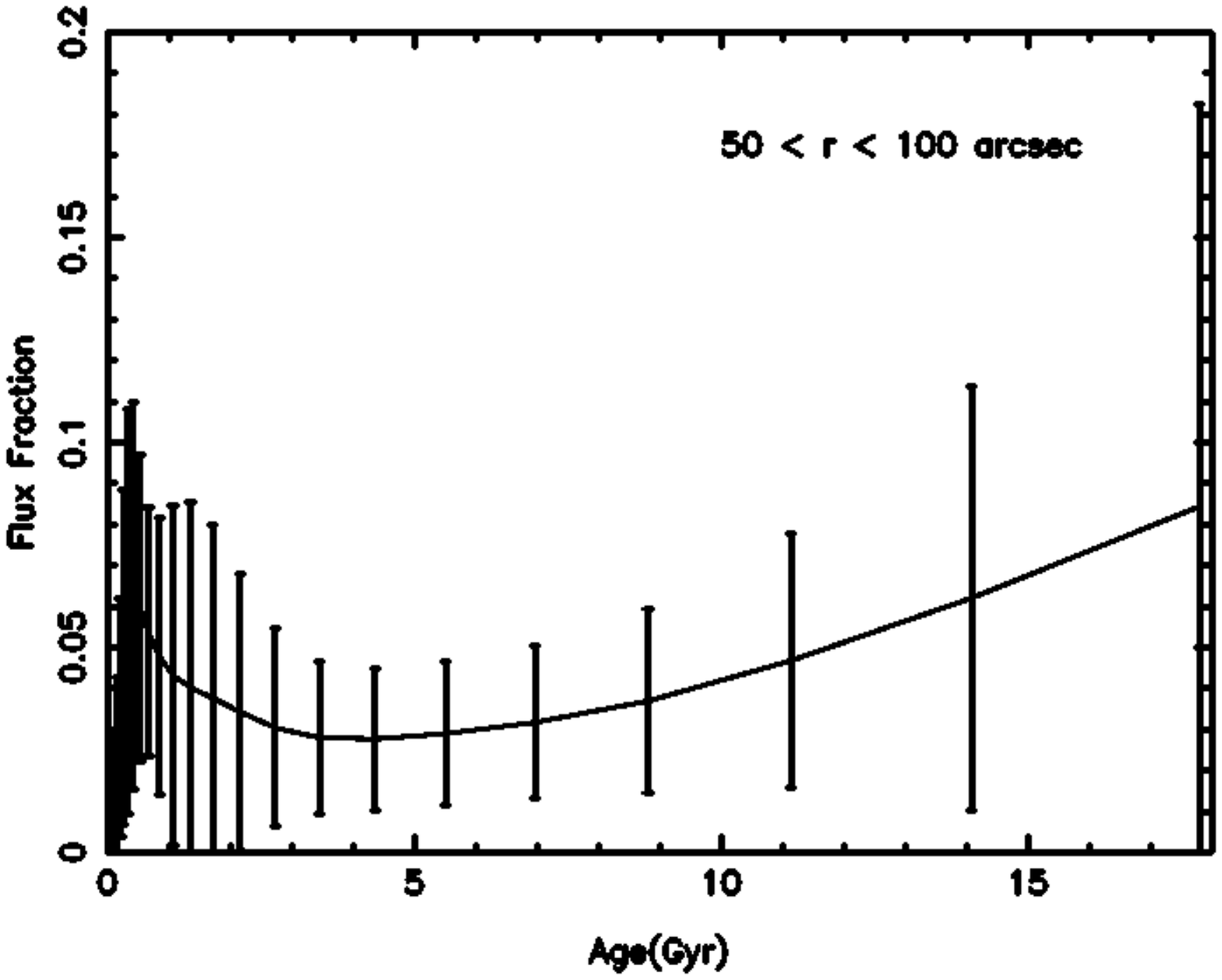}}
\resizebox{0.3\textwidth}{!}{\includegraphics[angle=-90]{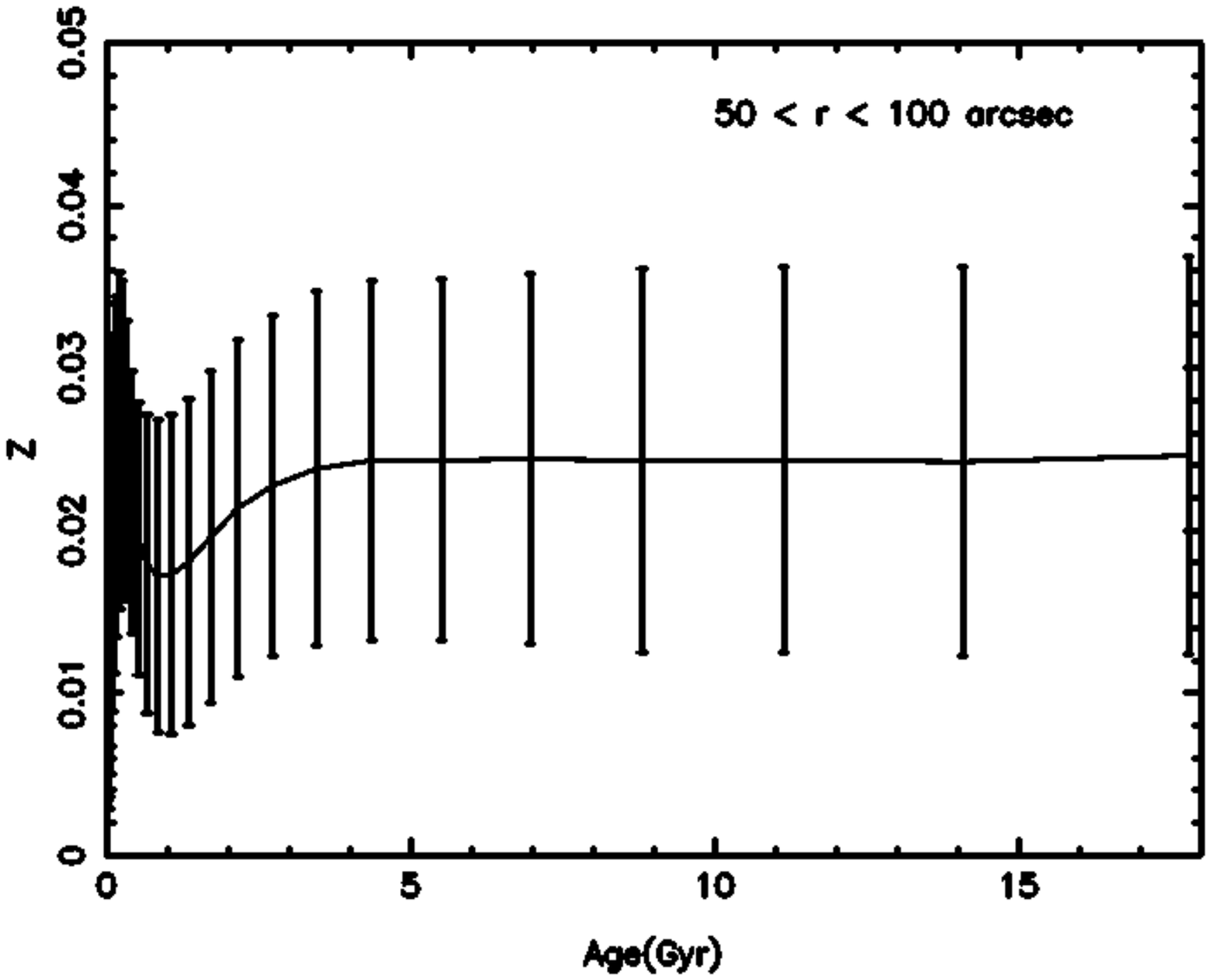}}\\
\caption{Star formation history and age-metallicity relation derived with {\tt STECKMAP} for the central 
spaxel and for all spectra
with radius between those indicated in the insets.
 The error bars represent the RMS scatter for all the averaged star formation histories\label{fig_sfr1}}
\end{figure*}

Figure~\ref{fig_sfr1} shows the mean flux fraction as a function of age  and the age-metallicity relation 
calculated at different distances from the center of the galaxy.
  It can be seen that, in the center, there is a younger population ($\sim$2 Gyr) that dominates the flux while its
contribution gets diluted as we go to larger radii.
We also see an increase in the metallicity with time, as it will be expected in a  normal 
chemical evolution model. 

\begin{figure*}
\resizebox{0.4\textwidth}{!}{\includegraphics[angle=-90]{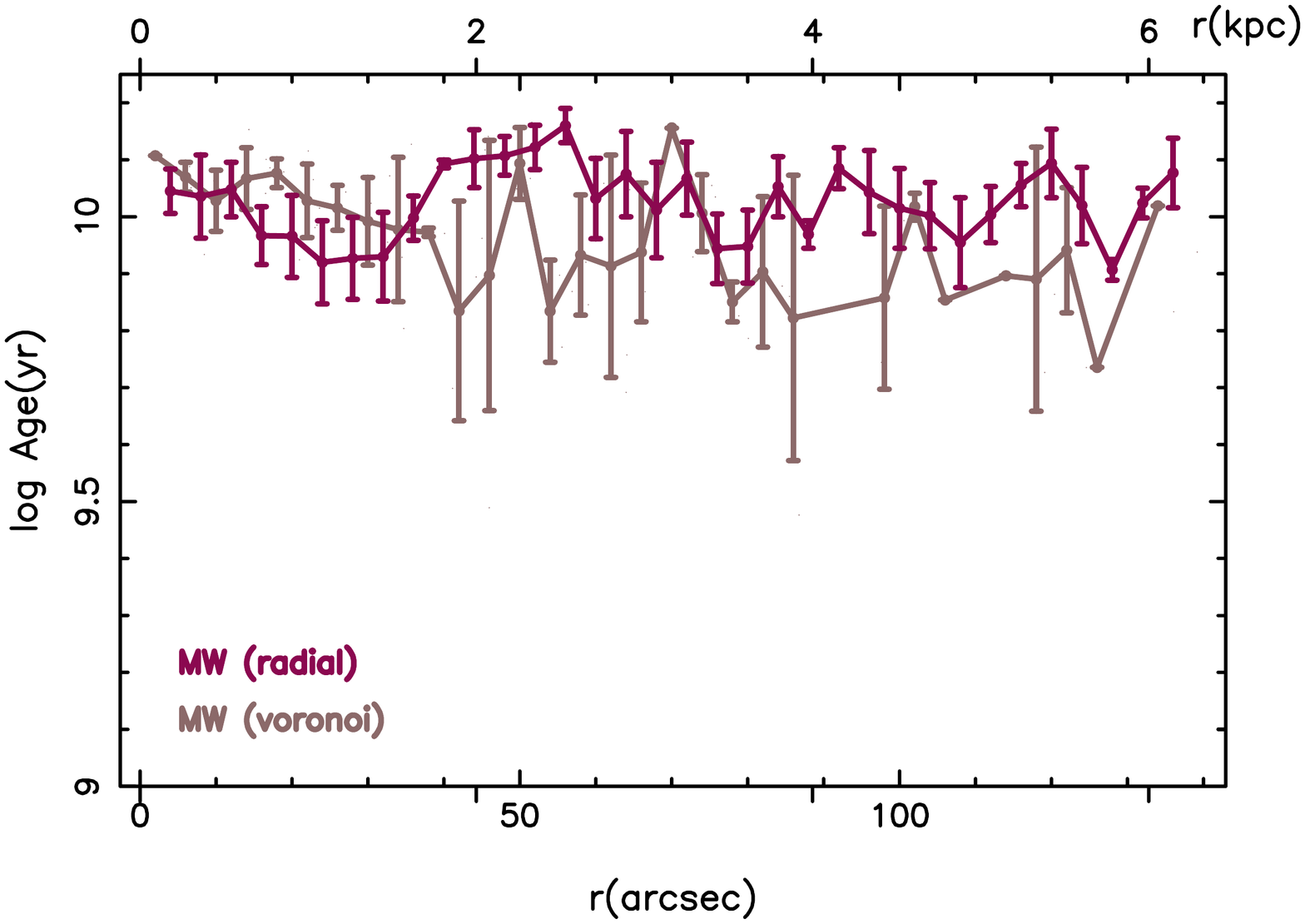}}
\resizebox{0.4\textwidth}{!}{\includegraphics[angle=-90]{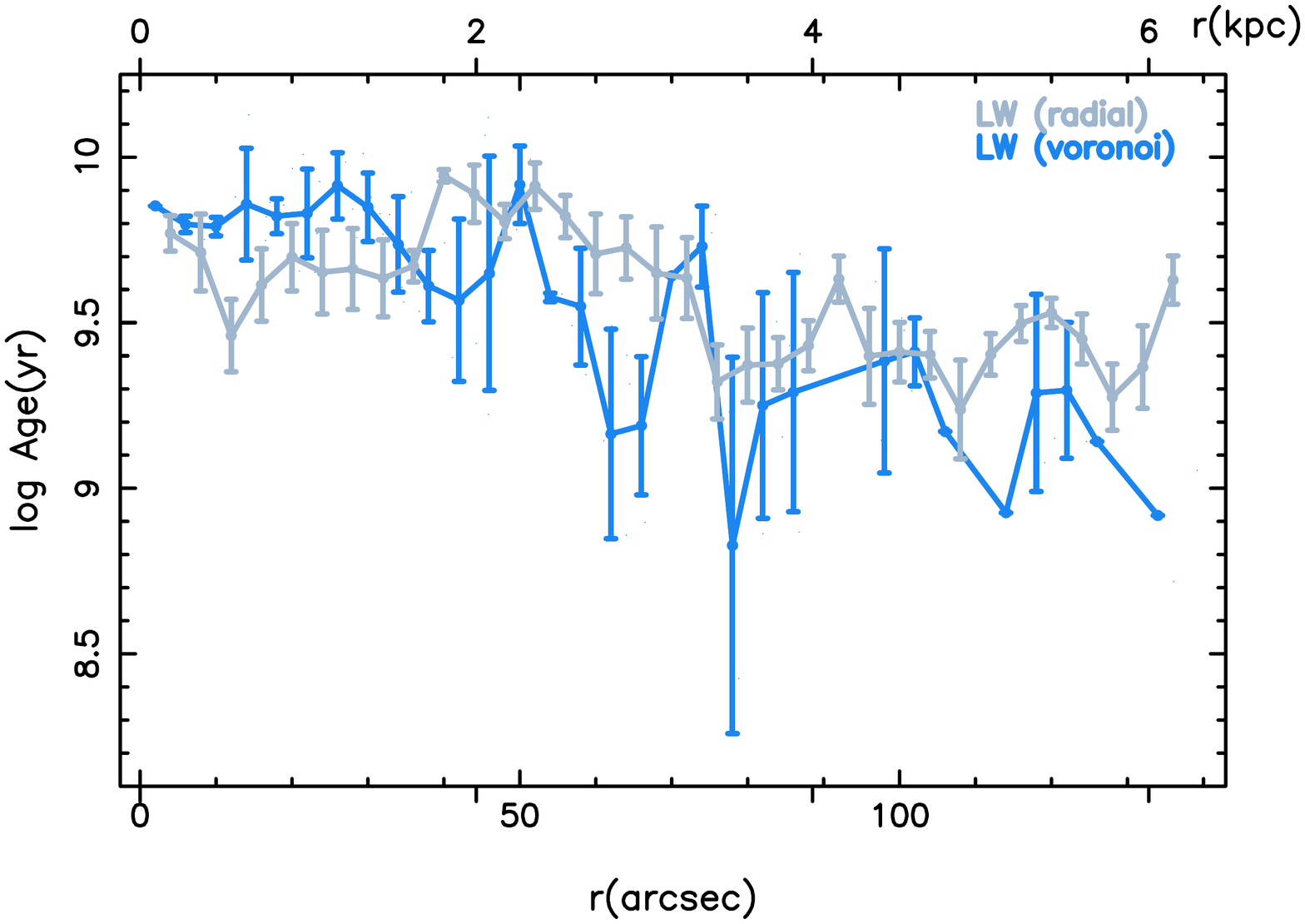}}
\resizebox{0.4\textwidth}{!}{\includegraphics[angle=0]{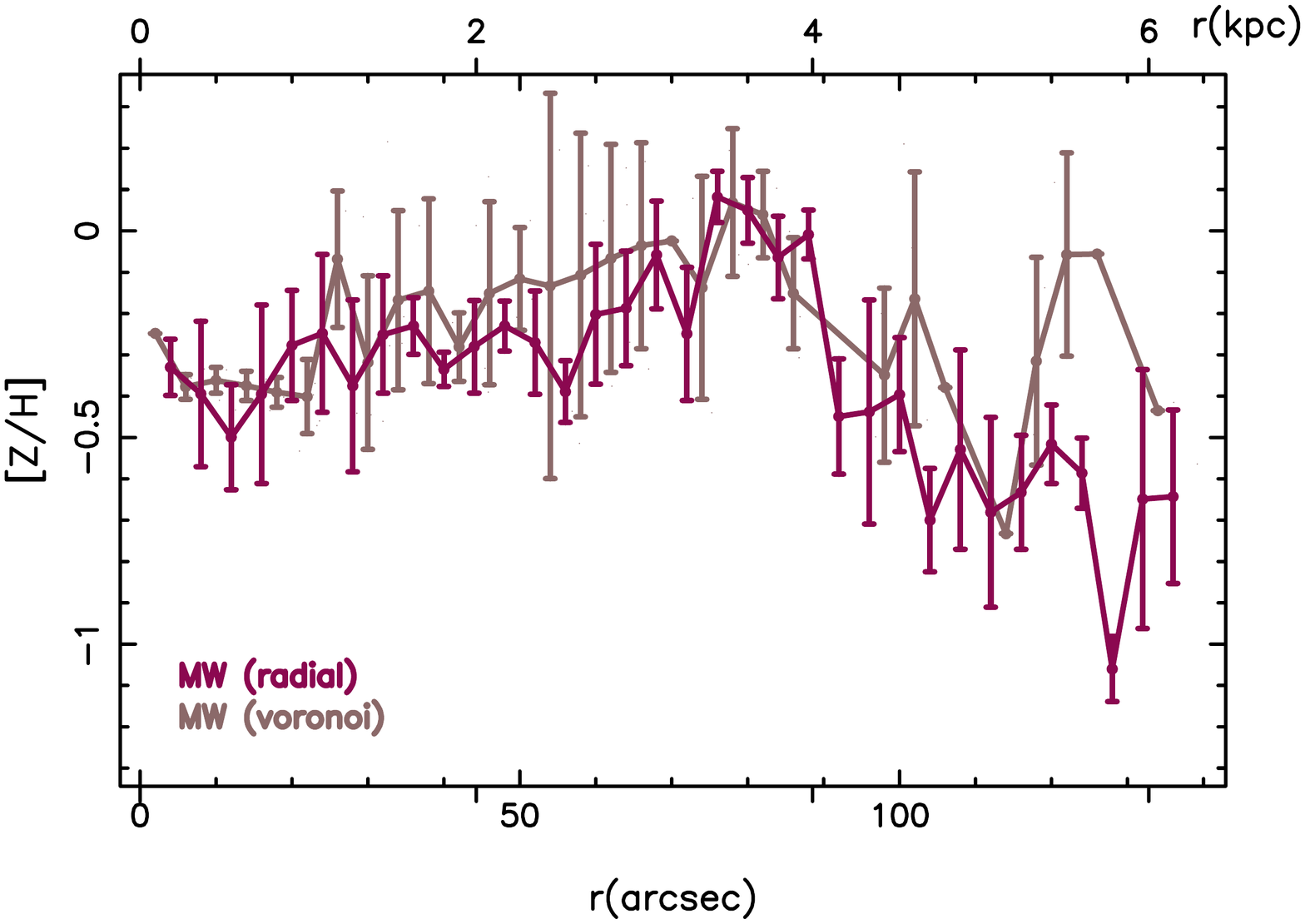}}
\resizebox{0.4\textwidth}{!}{\includegraphics[angle=0]{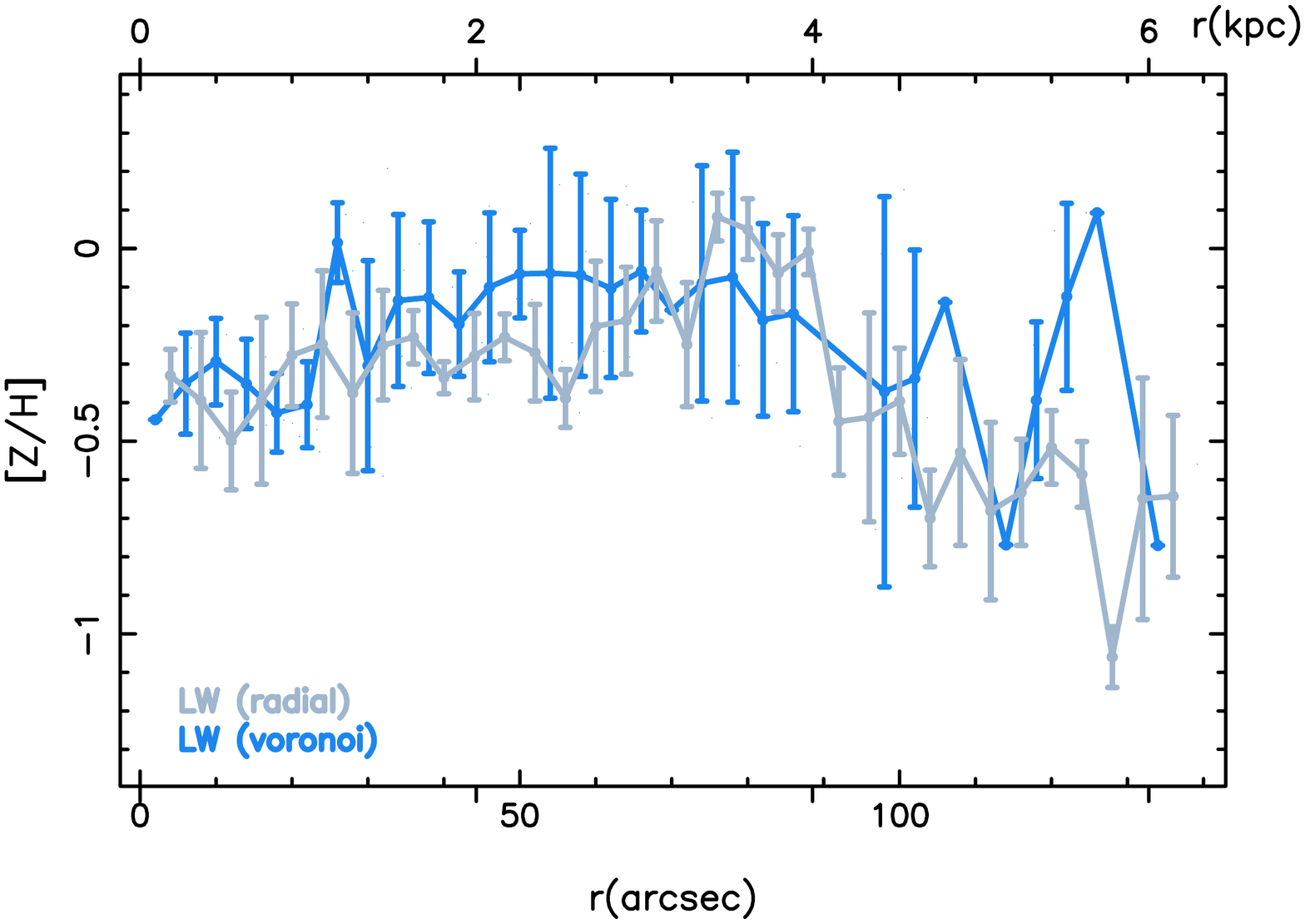}}
\caption{\label{gradients}Luminosity- (blue lines) and mass-weighted (red) age and metallicity gradients obtained 
for azimuthally averaged bins (radial) and averaged from the Voronoi bins (Voronoi). In the case of the Voronoi
binned values, the errors bars show the root-mean square of all the bins in each annulus.}
\end{figure*}

To quantify these trends we show, in 
Figure~\ref{gradients}, the mass and luminosity-weighted age and metallicity
gradients for NGC~628.

In both, age and metallicity profiles, there seems to be change of slope at around $\sim$60 and $\sim$80 arcsec
respectively.
We will analyse the gradients inside and outside this region separately.
\subsection{Age gradient}

The mass-weighted age of this galaxy is old ($\sim$10 Gyr)
at all sampled radii, i.e., the disc is dominated in mass by old stars, even at  $\sim$2 scale-length.
This is in agreement with S\'anchez-Bl\'azquez et al. (2011), where it was found that the mean, mass-weighted 
age of 4 disc galaxies was also around $\sim$~10~Gyr, at radii larger than 2~scale-length of the discs (see also MacArthur et al.\ 2009\nocite{MacArthurGonzalez&Courteau2009}).   

The luminosity-weighted mean age is biased towards the age of the youngest components, as these components are 
much more luminous in the wavelength range we are studying.
Here, it can be seen, as we have already mentioned, 
that the age gradient show two different behavior. In the inner
disc the gradient is  flatter (or even positive, with age increasing with radius).
A minimum in the mean luminosity-weighted age can also be seen in 
at $\sim$10$"$ (this is more readily seen in the azimuthally weighted values, where the errors are smaller), coincident with the radius of a circumnuclear star forming ring previously detected 
in sub-mm CO (1-0) observations (Wakker \& Adler 1995\nocite{Wakker&Adler1995}) and infrared 2.3 $\mu$m CO absorption
(James \& Seigar 1999)\nocite{James&Seigar1999} and was also seen by Ganda et al. (2006)\nocite{Gandaetal.2006} 
in their H$\beta$ and [OIII] maps of SAURON. 

In the outer disc, the luminosity-weighted age shows a clear gradient, with a variation of almost $\sim$6 Gyr in 2 scale-lengths. 

Previous studies of the stellar population across the radius for this galaxy include those of Natali et al.\~(1992)
\nocite{Natalietal.1992}\nocite{Zouetal.2011} and 
Zou et al. (2011) --using photometry in different bands -- and that of S\'anchez et al.\~(2011), using spectroscopy.
Zou et al. (2011) found, as in the present study, a different two distinct disc components, on from 30 to 60 arcsec, 
and the second one from 60 to 132 arcsec. At 1$'$, the age profile change slope. 
We also find a  change of slope at a similar radius but 
the trends that we derive are 
very different in both studies. Zou et al. (2011) found that the inner region of the disc (from 30-60$"$) have a much steeper
gradient  than the outer region, which is completely the opposite to what we find in the present study.
Our profile, however, coincides very  well with that of S\'anchez et al.\ (2011) (see Fig.~\ref{fig:sanchez}).

\subsection{Metallicity gradient}
The metallicity gradient also shows two differentiated regions,  being almost flat or slightly positive in the central 
parts and negative outwards.
The radius at which the metallicity gradient changes slope  
is  similar  to the break calculated by Scarano \& L\'epine (2013)\nocite{SL13} for the 
gas-phase metallicity using data from 
Rosales-Ortega et al.\ (2011),  and also similar to the break in the stellar metallicity found by Zou et al.\ (2011).
In fact, qualitatively, the trends found by Zou et al. (2011) are very similar to our; in the disc region the 
metallicity increase with radius until $R\sim60"$ and decrease slightly beyond. Our gradient in the internal regions is, however, much 
milder and the overall metallicities considerably larger those that derived by the quoted authors and the radius at which the gradient
change slope is larger.
\footnote{These authors adopted a different distance for this galaxy and, therefore, the radius
coincides in arcsec but not kpc.} 
This dual behavior  is, however,  not reported  by S\'anchez et al. (2011), although their luminosity-weighted values of age
and metallicity are in a very good agreement 
with ours (see Fig.~\ref{fig:sanchez}). 

\begin{figure}
\resizebox{0.5\textwidth}{!}{\includegraphics[angle=-90]{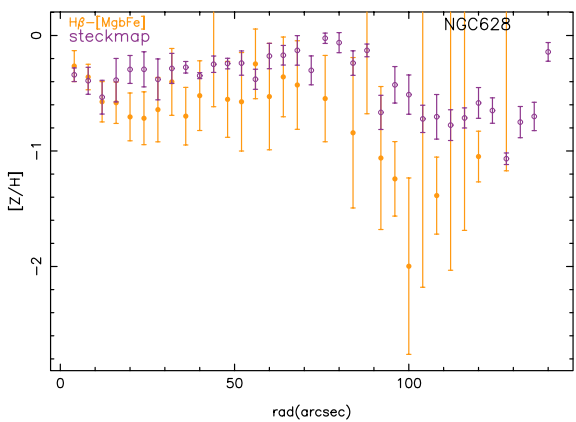}}
\caption{Comparison of the luminosity-weighted metallicity obtained with {\tt STECKMAP} (purple, solid circles) and 
the single-stellar population equivalent one obtained in a diagram combining H$\beta$ and [MgbFe]. \label{compara}}
\end{figure}

To check the reliability of this break  we compare  the metallicities obtained with our methods
with those obtained with a better understood and widely used method, a classical index-index diagram. Figure~\ref{compara} shows the comparison of the metallicity gradient
obtained with {\tt STECKMAP} and the one obtained using an index-index diagram combining H$\beta$ and [MgbFe].  We are not 
comparing exactly the same quantity as the metallicity obtained by using an index-index diagram is a single-stellar population equivalent
value, while the one obtained  with {\tt STECKMAP} is a luminosity-weighted one. However, it has been shown that these two measurements give very 
similar values (Serra \& Trager 2007)\nocite{ST07}.  It can be seen that, although the SSP-equivalent values are noisier,
the trends obtained with 
the two methods are very similar and, in particular, a change of slope is visible  at the same radius.

The different trends in age and metallicity between this study and that of Zou et al.\ (2011) shows the difficultly in 
deriving accurate stellar population parameters using broad band colors, due to the existent degeneracies between age, metallicity 
and dust extinction. It is worth noticing that the broad band optical color  profiles of this galaxy show, in fact, a very steep slope in the inner 
region of the disc while it flattens out beyond 1$'$. On the contrary, we obtain much flatter colors of age and metallicity in the 
inner regions of the disc. We prove, however, in Sec.~\ref{comp:photometry} that we can reproduce the observed colors of this galaxy using our 
derived star formation histories. 

\subsection{Time evolution of the gradients}
One of the advantages of being able to derive an age-metallicity relation is that
it allows us, in principle, to obtain the time evolution of the metallicity gradient.
The temporal evolution of the metallicity gradient is an issue that is not yet settled from neither, the 
theoretical or the observational point of view.
Some chemical evolution models predict a steepening with time starting from initially inverted or flat
gradients (e.g. Chiappini, Matteucci \& Romano 2001\nocite{2001ApJ...554.1044C}) while others predict an initially negative gradient that
flattens (e.g., Moll\'a \& D\'{\i}az 2005\nocite{2005MNRAS.358..521M}).
Recent work (Pilkington et al.\ 2012\nocite{2012A&A...540A..56P}; Gibson et al.\ 2013\nocite{2013A&A...554A..47G})
have showed that, in cosmological simulations, the evolution of the metallicity gradient
with time depends on the details of the sub-grid physics implemented in the
hydrodynamical codes of galaxy formation, in particular, on the feedback scheme. 
Therefore, the study of the metallicity gradient for
different age population in galaxies are key to reveal the formation processes of disc galaxies
and better constrain the physics included in numerical simulations.

Some authors have studied this evolution in our own MW using  
planetary nebulae (e.g., Maciel et al. 2003\nocite{2003A&A...397..667M}; Stanghellini \& Haywood 2010\nocite{2010ApJ...714.1096S}; 
Maciel \& Costa 2013\nocite{2013arXiv1308.1884M}) also with discrepant results.
A direct measurement of the evolution of the metallicity gradients can be obtained analysing  
the magnitude of the gas-phase metallicity gradient at high redshift (Cresci et al.
2010\nocite{2010Natur.467..811C}; Jones  et al.\ 2010\nocite{2010ApJ...725L.176J}; 
Yuan et al. 2011\nocite{2011ApJ...732L..14Y}; Queyrel et al. 2012\nocite{2012A&A...539A..93Q}). Yuan et al.\ (2011) found 
that, for at least one grand design spiral at redshift z~1.5, the metallicity gradient is
significantly steeper ($-0.16$ dex/kpc) than the typical gradient encountered today.
However, the difficulties in obtaining high resolution data for likely MW progenitors has mean
that the theoreticians have had very few constraints on their  models, as different paths can
lead to a similar final metallicity gradient. 

\begin{figure}
\centering
\resizebox{0.45\textwidth}{!}{\includegraphics[angle=0]{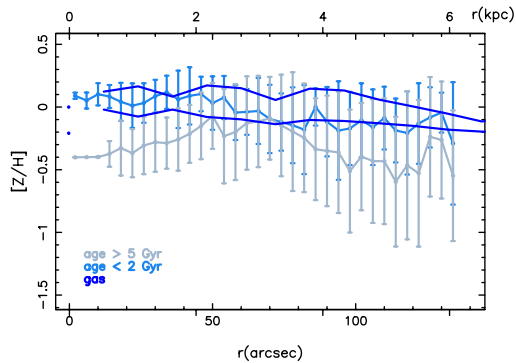}}
\caption{Mean (luminosity-weighted) metallicity considering only those 
populations with age younger than 2 Gyr  and older than 5 Gyr . Error
bars represent the RMS dispersion of the metallicity obtained in 25 simulations where each pixel 
of the spectrum is perturbed
according to its signal-to-noise following a Gaussian distribution. 
The dark blue line shows the gas-phase metallicity gradient obtained by Rosales-Ortega et al. (2011)
using two different calibrators, ff-T$_{\rm eff}$ and O3N2 (see the text for details). To convert
the scales we have adopted a solar abundance of 12+log(O/H)=8.7.\label{evol_grad}}
\end{figure}

Figure~\ref{evol_grad} shows the spatially resolved metallicity
in the HII regions of this galaxy (taken from Rosales-Ortega et al. 2011) and the stellar components younger than 2~Gyr and older than 5~Gyr.
To compare the gas- and stellar metallicities we have transformed the
former to the scale of the later assuming
a solar metallicity of 12+log(O/H)=8.7 (Asplund et al.\ 2005)\nocite{Asplundetal2005}.
It can be seen that, if we restrict ourselves to the younger stars (those with ages$<$2 Gyr), the
gradient shows a linear behavior, decreasing slightly with radius, and it is very similar 
to the gradient of the ionized gas.
On the contrary, the metallicity profile for the older stars
have a different slope inside and outside $\sim$50-60$''$. 
The metallicity gradient for the older populations is slightly positive
in the internal zones, but it is steeper than that of the young population in the external 
parts. 

Note that, although 
in Paper~I and in Scarano \& L\'epine (2013)\nocite{SL13} a change of slope in the gas-phase metallicity --
at about the same radius as the one found here for the stellar component -- is reported, 
this  is much more subtle than the one observed in the metallicity gradient of the old stars (Fig.~\ref{evol_grad})\footnote{ 
The line represents the values obtained by Rosales-Ortega et al. (2011), not those of Scarano \& L\'epine (2013), 
although the data used by both studies are the same}.
\section{Comparison with other methods and robustness of the solution}
\label{sec:comparison}

\subsection{Comparison with S\'anchez et al. 2011}
\label{comparison_sebas}
\begin{figure*}
\resizebox{0.4\textwidth}{!}{\includegraphics[angle=0]{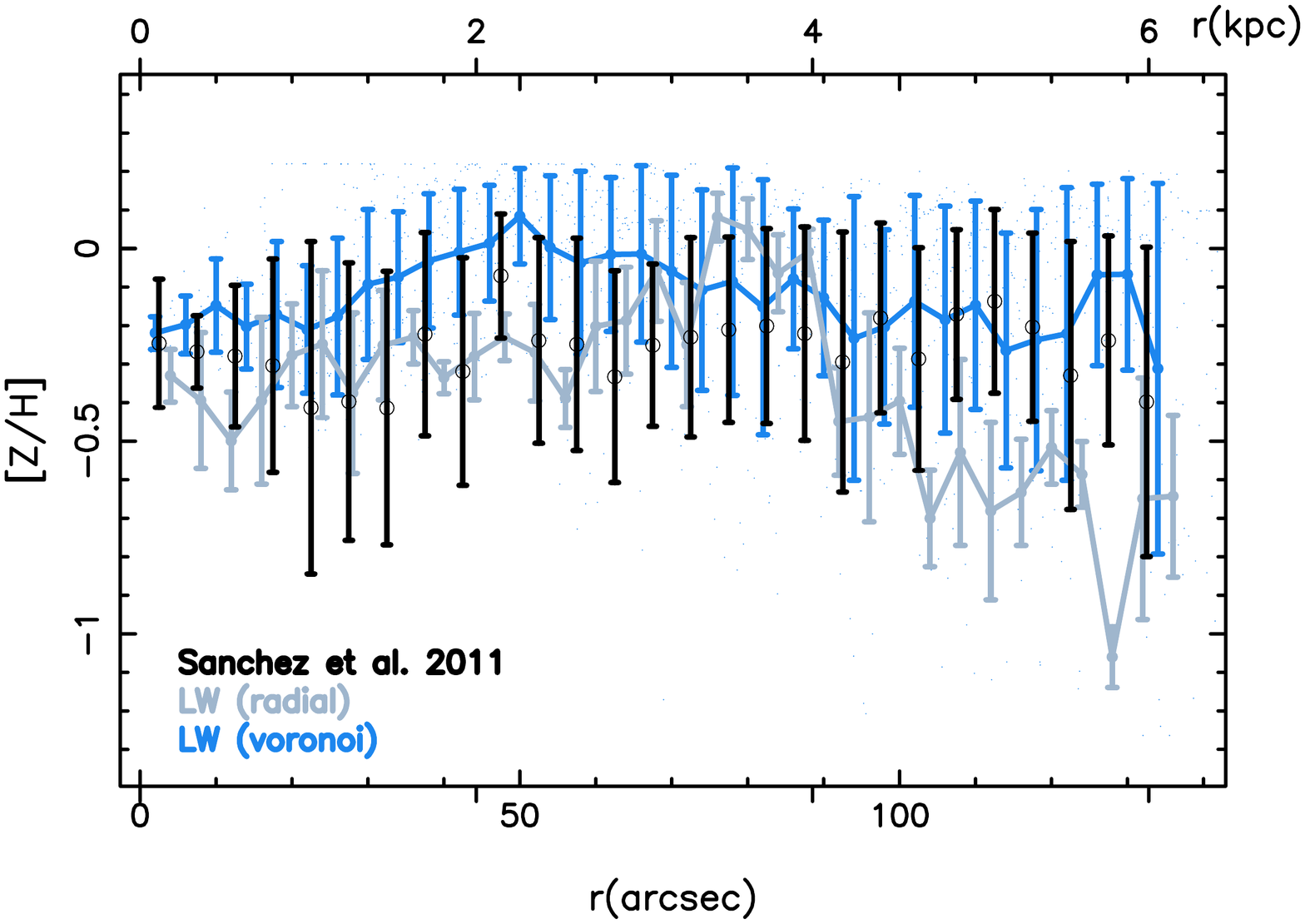}}
\resizebox{0.4\textwidth}{!}{\includegraphics[angle=0]{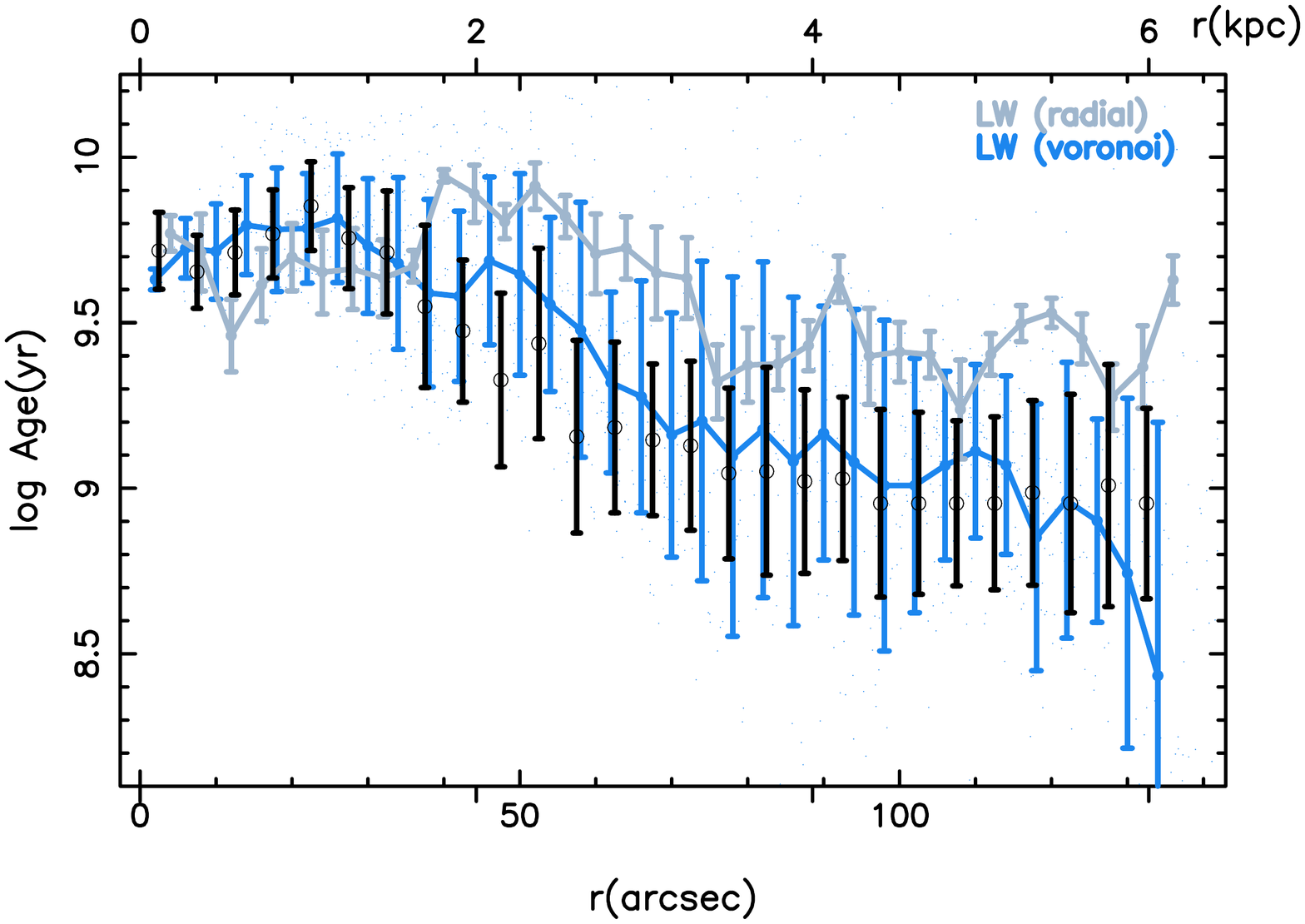}}
\caption{Comparison of the luminosity-weighted gradients of age and metallicity between 
S\'anchez et al. (2011) and the present study.\label{fig:sanchez}}
\end{figure*}

S\'anchez et al. (2011) presented an analysis of the stellar population properties
for the same galaxy using also the same data as the one presented here. These authors
used {\tt FIT3D} (S\'anchez et al. 2007bc). {\tt FIT3D} enables linear fits of a combination 
of single stellar populations, and non-linear ones of emission-lines plus an underlying stellar 
population. It also includes the Cardelli, Clayton \& Mathis (1989)\nocite{1989ApJ...345..245C} attenuation law (see S\'anchez
et al.\ 2011 for details).
In S\'anchez et al.\ (2011), {\tt FIT3D} was used in combination with Bruzual \& Charlot (2003)\nocite{Bruzual&Charlot2003} models, 
assuming a Salpeter IMF. The comparison between their results and those of the present study are 
shown in Fig.~\ref{fig:sanchez}. It is remarkable the excellent agreement between both studies, despite
the differences in the code and in the stellar population models used. This agreement gives particular confidence 
in the results.

\subsection{Comparison with the photometry}
\label{comp:photometry}
\begin{figure}
\resizebox{0.33\textwidth}{!}{\includegraphics[angle=-90]{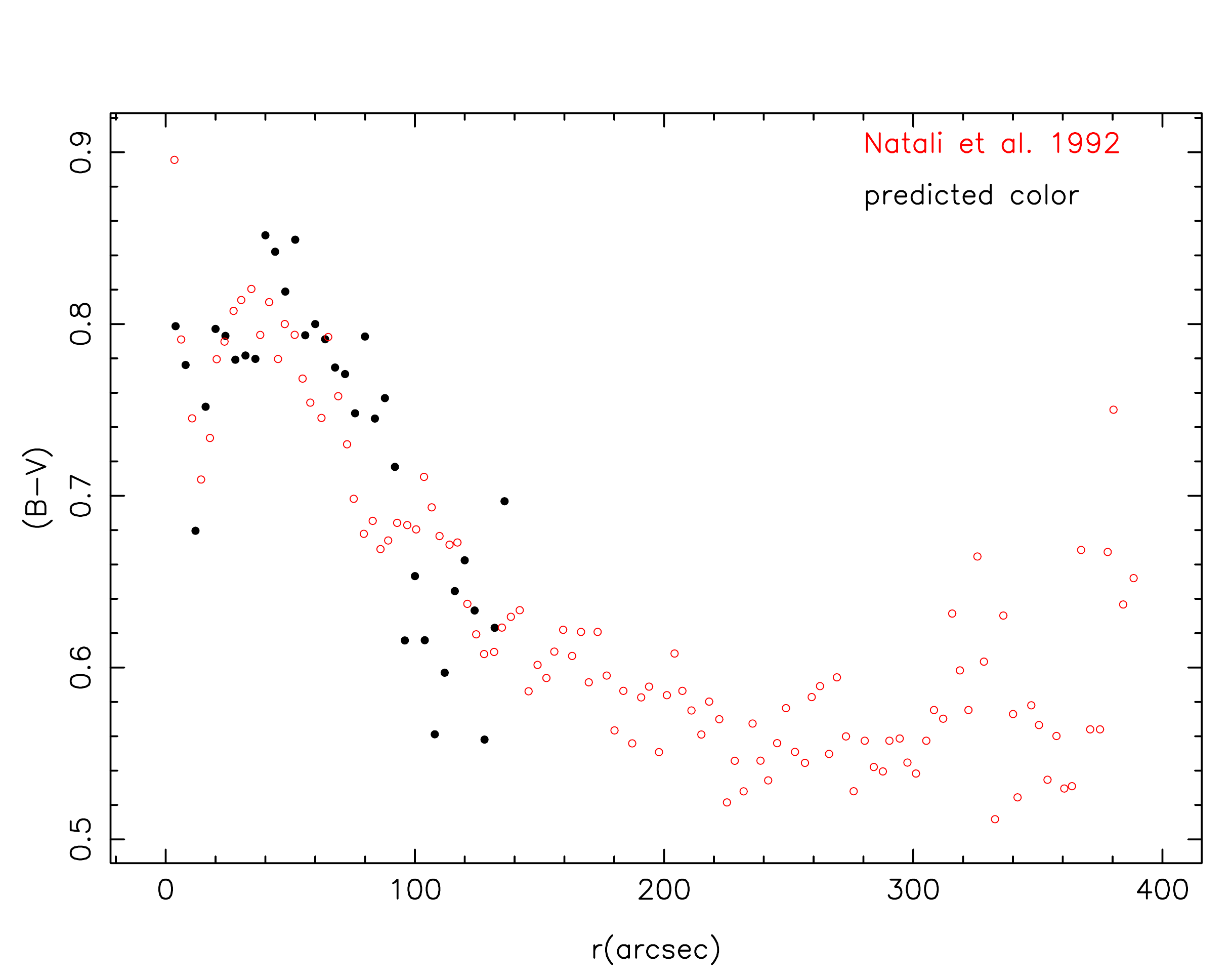}}
\resizebox{0.35\textwidth}{!}{\includegraphics[angle=0]{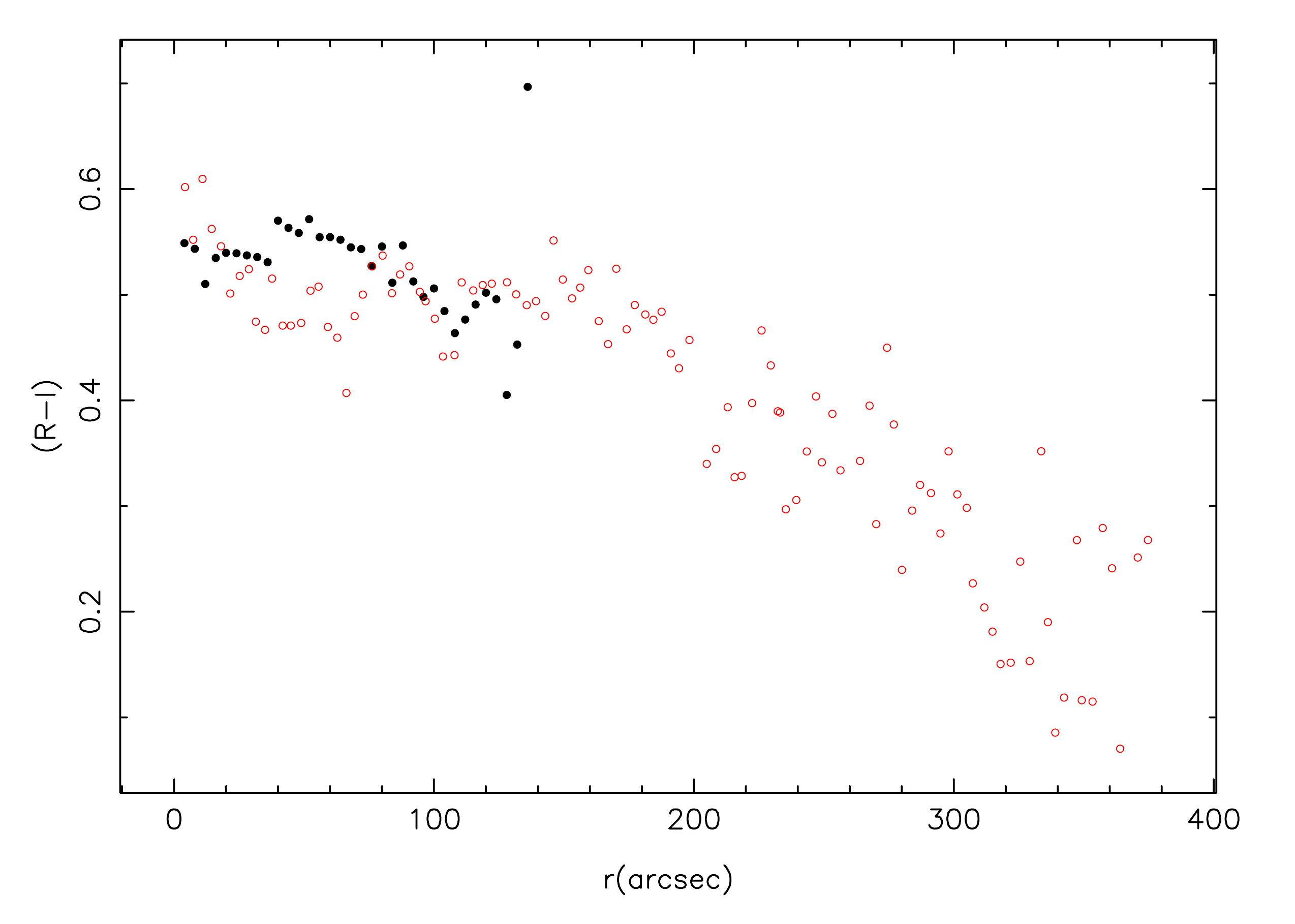}}
\caption{Comparison of the R-I and B-V color gradients derived by Natali et al.\ (1992)
(red open circles) and those derived by us from the star formation histories and age-metallicity relation (black circles).
\label{color_predicted}}
\end{figure}

NGC~628 has been extensively studied by many authors. We can take advantage of the available 
data to check the methodology presented in this paper. Natali et al. (1992)\nocite{Natalietal.1992} performed surface photometry 
of this galaxy and presented U-V and B-V colors along the radius corrected by extinction. One way to test the viability 
of our employed inversion technique is to use our derived star formation histories and age-metallicity 
relation to infer a colour gradient, by adding the colour corresponding to each population weighting with 
the luminosity in the corresponding photometric band. 
We did this, and the results are shown in Fig.~\ref{color_predicted}.
As can be seen, the agreement is very good in the radial region covered by our data. 

\section{Discussion}
\label{sec_discussion}
It is clear from previous studies that NGC~628 is a perfect candidate for a galaxy having experienced 
secular evolution. Its bulge has the characteristic of a bulge formed secularly (Kormendy \& Kennicutt 2004\nocite{KK04}),
with a surface brightness profile showing a S\'ersic index $<$2 (Zou et al. 2011\nocite{Zouetal.2011}), presence of nuclear
spiral arms (Cornett et al. 1994\nocite{Cornettetal.1994}) and a young ring structure, a nuclear velocity dispersion 
drop (Ganda et al. 2006\nocite{Gandaetal.2006}). Fathi et al. (2007)\nocite{Fathietal.2007} found indications
that the gas is falling in from the outer parts towards the central regions, where a nuclear ring has formed 
at the location of the inner Lindblad resonance radius of an m=2 perturbation.
We have analysed the stellar population gradients in the galaxy NGC~628 and found that the stellar population 
gradients in the disc shows a two distinct regions, one older and more metal rich with almost flat gradient 
in metallicity and slightly positive in the mean (luminosity-weighted) age and an external part with slightly decreasing gradients 
in both, luminosity-weighted age and metallicity. The mass-weighted age gradient is compatible with being flat in all the regions
sampled by our data, indicating that this galaxy is dominated, in mass, by old stars ($>$9~Gyr).

\subsection{The outer disc}
Outside r$\sim$1$'$, 
the analysis above shows that the majority of the stars in the disc of this  galaxy 
are old, even at $\sim$ 2 scale-lengths of the disc, but that the proportion 
of young stars is higher at larger radii, i.e., the outer parts of the disk formed a larger 
fraction of their stars at recent times than the inner parts, consistent
with the $"$inside-out$"$ growth scenario. However, even the outer regions
sampled by us, contain a large percentage of old stars, reflected in an old mass-weighted mean age. 
Studies of stellar populations using resolved stars have found old stellar populations 
in the outskirts of disc galaxies (e.g., Ferguson \& Johnson 2001 (M31)\nocite{2001ApJ...559L..13F}; 
Davidge 2003\nocite{2003AJ....125.3046D} (NGC2403, M33); Galleti, Bellazzini \& Ferraro 2004\nocite{2004A&A...423..925G} (M33); 
Gogarten et al. (2010)\nocite{2010ApJ...712..858G} 
(NGC300)). It has been suggested that these properties are not expected in 
CDM models (Ferguson \& Johnson 2001\nocite{2001ApJ...559L..13F}) where the disc grows inside-out and at 
relatively recent epochs ($z \leq$ 1). However, in S\'anchez-Bl\'azquez 
et al. (2009)\nocite{Sanchez-Blazquezetal2009}, we analysed a fully cosmological, hydrodynamical disc
galaxy simulation in a $\Lambda$CDM Universe finding, 
also, a large percentage of  old stars in the outskirts. 

On the other hand, numerical simulations have shown that  radial migrations 
due to the presence of non-axisymmetric components, like bars or spiral arms, 
can make that an important percentage of stars currently in the outer parts
of the disc may have formed closer to the center of the disc and migrated
outward to their current locations. Therefore, an observed old stellar 
population in the outer disc may not necessarily indicate that the outer
disc formed early. 
In the present study, we have been able to analyse, for the first time, the evolution
of the stellar metallicity gradient with time from integrated spectra. In the $"$outer$"$-disc
region, we have found that the metallicity gradient for the old stellar population
is steeper than that of the young component. 
This, in principle, can be a sign that stellar migration has not been very important
in the disc region of this galaxy. However, 
to conclude this we would need to know the original {\it gas-phase} metallicity gradient 
at the epoch of formation of these old stars, which could have been even steeper.

\subsection{The inner disc}

At radius $R <\sim 60''$  both, the mean age and metallicity profiles, are flatter than in the external disc.
In fact, the gradient is even positive for the luminosity-weighted values of age and  metallicities. 
This is difficult to explain in an inside-out 
formation 
scenario for the disc formation and requires an explanation.
The change in the slope of the gas-phase metallicity gradient has
been observed
before in many galaxies (e.g., Zaritsky 1992\nocite{Zaritsky92}; Vila-Costas \& Edmunds 1992\nocite{VE92};
Martin \& Roy 1995\nocite{MR95};
Dutil \& Roy 1999\nocite{DR99}; Roy \& Walsh 1997\nocite{RW97};
 Zahid \& Bresolin 2011\nocite{ZB11}; Scarano \& L\'epine 2013\nocite{SL13})
 and it is usually interpreted as the  result of variations of gas density,
 the large-scale mixing induced by a bar
 (Friedli \& Benz 1995\nocite{FB95}), or by the spiral arms. In this case, 
 the change of slope coincide  with the corotation radius of the spiral patterns (e.g.,
 Scarano, L\'epine \& Marcon-Uchida 2011\nocite{SLM11}; Scarano \& L\'epine 2013\nocite{SL13}).
 This is because the main gas flows have opposite directions inside and outside  corotations.
 However, the above mechanism applies to young objects, such HII regions, that
 reveal the present value of the metallicity in the interstellar medium
 The fact that the break in the stellar metallicitiy is more pronounced in the older component points toward
 dynamical effects (radial migrations) as the cause for the observed dual behavior.

Flat gradients, as the ones found here in the inner disc-- require strong radial mixing, which is difficult to 
produce with star formation histories but can be achieved in ongoing strong galaxy interactions (Rupke et al. 2010\nocite{2010ApJ...710L.156R})
or can be produced by secular evolution in the presence of non-axisymmetric components, like bars, oval or spiral arms. 
NGC~628 is classified as an unbarred galaxy and, therefore, if secular evolution has happened, one is tempted to attribute
it to the presence of spiral arms. 
Strong resonant interaction with transient or long-lived spiral arms at the corotation radius can produce
stellar wanderings of several kpc in a few hundred Myr (e.g., Sellwood \& Binney 2002\nocite{SB02}; 
L\'epine, Acharova \& Mishurov 2003\nocite{2003ApJ...589..210L}; Ro{\v s}kar et al. 2008a\nocite{2008ApJ...675L..65R}).
If the spirals are long-lived, this mechanism has been shown to produce bimodal patterns in the metallicity gradients with 
breaks at the spiral corotation radius.  The corotation radius for NGC~628 is located around $\sim$7~kpc 
(Sakhibov \& Smirnov 2004\nocite{2004ARep...48..995S}) which is outside the region sampled by us.
Therefore, the break that we observe here is too internal to be produced by the spiral arms.
However, a change on the slope of the color gradients around this radius has been reported in Natali et al.\ (1992).

On the other hand, despite  NGC~628 is classified as an unbarred galaxy in de Vaucouleurs et al.\ 1991 atlas),
near-infrared imaging and isophotal analysis  has revealed the existence of an oval distortion with radius of r$\sim$50$''$ 
(Laine et al. 2002; Seigar 2002\nocite{Seigar2002}) confirmed by the kinematical analysis by Fathi et al. (2007)\nocite{Fathietal.2007}.
These authors also conclude that this oval (a m=2 perturbation) is the most prominent in the observed
velocity field.
Thought ovals are rather frequent in disk galaxies, their exact nature has not yet been investigated. Kormendy (1979)\nocite{1979ApJ...227..714K}
hypothesized that they are products of bars decays, which is supported by some observations of galaxies
with prominent rings (e.g. Sil'chenko \& Afanasiev 2002\nocite{2002A&A...385....1S}).
 Such a scenario is also  predicted by the models of Friedli \&
  Benz (1993, 1995\nocite{FB95}\nocite{1993A&A...268...65F}); Combes et al. 1990\nocite{1990A&A...233...82C}; 
Raha et al. 1991\nocite{1991Natur.352..411R}; Martinez-Valpuesta \& Shlosman 2004\nocite{2004ApJ...613L..29M}.
The presence of a circumnuclear ring of star formation (Wakker \& Adler 1995\nocite{Wakker&Adler1995};
 James \& Seigar 1999\nocite{James&Seigar1999}) and a central 
$\sigma$-drop (Ganda et al. 2006) in NGC~628 support this hypothesis, as the large-scale oval perturbation 
can transport material from the outer zones to regions near the center of the galaxies (see Fathi et al. 2007 for discussion).

Our results also show that we have found that the stellar population in the region dominated by the oval distortion 
reveals a positive luminosity-weighted age and metallicity gradients.
This has been observed in bars (P\'erez, S\'anchez-Bl\'azquez \& Zurita 2009\nocite{2009A&A...495..775P}) and, in fact, 
Moorthey \& Holtzman (2006)\nocite{2006MNRAS.371..583M}, found that the only galaxies with  positive metallicity gradients
in the central regions were those hosting bars. 
Positive stellar population gradients have also been reproduced in chemodynamical simulations 
of barred galaxies (Wozniak et al.\ 2007\nocite{2007A&A...465L...1W}).

Weak oval structures are not believed to have a strong influence in the redistribution of stars inside the galaxy.
However, this structures could have been stronger in the past. In fact, some authors
(Kormendy 1979\nocite{1979ApJ...227..714K}; 1981\nocite{1981seng.proc...85K}, 1982\nocite{1982ApJ...257...75K}; Combes 1996\nocite{1996ASPC...91..286C}) 
suggest that 
some bars evolve into lens components.
Therefore, the arguments above  suggest  that we could be  witnessing a bar in the process of being destroyed, support coming from 
previous kinematical studies and from our stellar population analysis. 

Understanding bar formation and evolution is crucial to understand galaxy evolution, yet a fundamental 
question, namely whether bars can be destroyed by internal processes, remains unanswered.
From the theoretical point of view, there is currently a debate in the literature. On one hand, bars
seem to be robust structures  which cannot be easily destroyed by central mass concentration as
the observed are not massive enough (Shen \& Sellwood 2004\nocite{2004ApJ...604..614S}; Athanassoula, Lambert \& Dehnen 2005\nocite{2005MNRAS.363..496A}). 
On the other hand
some numerical simulations predict that bars can be destroyed and reformed when disc are very gaseous, and 
that galaxies can have up to 3-4 generations of bars over a Hubble time (Bournaud \& Combes 2002\nocite{2002A&A...392...83B}).
Recent observations are consistent with the picture in which bars in early-type disc galaxies are long-lived, 
whereas those in late-type spirals could be dynamically young (Gadotti \& de Souza 2005\nocite{2005ApJ...629..797G}, 2006\nocite{2006ApJS..163..270G}; 
Elmegreen et al. 2007\nocite{2007iuse.book..163E};
Sheth et al. 2008\nocite{2008ApJ...675.1141S}). However, there is no observational result to date that indicates whether bars can dissolve.
In particular, we cannot tell whether an unbarred galaxy has had a bar in the past. However, ovals (or lenses)
can give us some clues, as one of the proposed formation scenarios is that lenses could be the remnants of a dissolved bar.
  Here we show stellar population gradients that are compatible with this 
scenario as, otherwise, it would be  difficult to obtain the change on the slope of the stellar population parameters at the radius we are observing them and 
the slightly positive metallicity  gradient in the inner disc.

\section{Conclusions}
\label{sec_conclussion}
In this paper we have presented an analysis of the star formation history and metallicity evolution  in a 2-dimensional
fashion for the galaxy NGC~628, using data form the survey PINGS (Rosales-Ortega et al. 2010) with the main purpose
of validate our techniques to derive reliable stellar population properties.
\begin{itemize}
\item We derive 2D-star formation histories for NGC~628 using the code {\tt STECKMAP}. The analysis 
allows to obtain mean ages and metallicities weighted with both, luminosity and stellar mass.
\item We check the robustness of the derived mean ages and metallicities to different choices of the fit. In particular,
we have checked the effect of deriving simultaneously kinematics and metallicities, and also the effect or 
fitting or not the stellar continuum. We have found that, when the kinematics is fitted simultaneously to the 
derivation of star formation histories, the ages are not very affected, but the derived metallicities are. 
We have
also found that the differences can be minimized in spectra with enough signal-to-noise.

\item We find a negative luminosity-age gradient compatible with an inside-out formation scenario 
(e.g., Larson 1976\nocite{1976MNRAS.176...31L}; White \& Frenk 1991\nocite{WF91}; Mo et al. 1998\nocite{1998MNRAS.295..319M}; 
Naab \& Ostriker 2006\nocite{2006MNRAS.366..899N}; Brook et al. 2006\nocite{2006ApJ...639..126B}).
However, even at $\sim$2.0 scale-lengths the disc (assuming a disc scale-length of 3.2 kpc -- Zou et al. 2011)
is dominated, in mass, by old stars ($\sim$10~Gyr).

\item We find a metallicity gradient that is flatter in the internal region and change to a negative slope 
in the external parts. 
The break in the metallicity gradient is much more prominent in the old stars than in the younger components, 
which gradient is very similar to the gas-phase metallicity gradient. However, a break is visible in all the different 
components at about the same radius. We speculate that the metallicity gradient is the consequence of the gaseous flows and 
stellar migration produced by a $m=2$ perturbation, either the spiral pattern of a bar. However, the position of the break is 
more coincident with the corotation radius of the oval distortion than that of the spiral pattern, which is beyond the 
radius sampled by our data.
As the oval distortion is NGC~628 is too weak for this to happen very 
efficiently, it could be  that there was a stronger bar present in this galaxy and that the oval perturbation is the consequence
of the dissolution of this structure. We also argue that this structure cannot have evolved much in size in, at least,  the last $\sim$
5 Gyr.
\end{itemize}

\section*{Acknowledgments}
This work has received support form the Spanish Ministerio de Ciencia e Innovaci\'on
under the research projects AYA2010-21322-C03-03 (RAVET)
and AYA2010-21887-C04-03 (ESTALLIDOS) and AYA2012-31935. PSB and SFS
acknowledge support from the Ram\'on y Cajal program of the Spanish Ministerio de Econom\'{\i}a y 
Competitividad.
FFRO acknowledges the Mexican National Council for Science and Technology (CONACYT) for financial
support under the programme Estancias Posdoctorales y Sab\'aticas al Extranjero para la Consolidaci\'on
de Grupos de Investigaci\'on, 2010-2011.
Based on observations collected at the Centro Astronómico Hispano-Ale\'an (CAHA) at Calar Alto,
operated jointly by the Max-Planck Institut f\"ur Astronomie and the Instituto de Astrof\'{\i}sica de
Andaluc\'{\i}a (CSIC).

\appendix

\section{Effect of fixing the kinematics}
\label{appendix2}
\begin{figure*}
\resizebox{0.6\textwidth}{!}{\includegraphics[angle=-90]{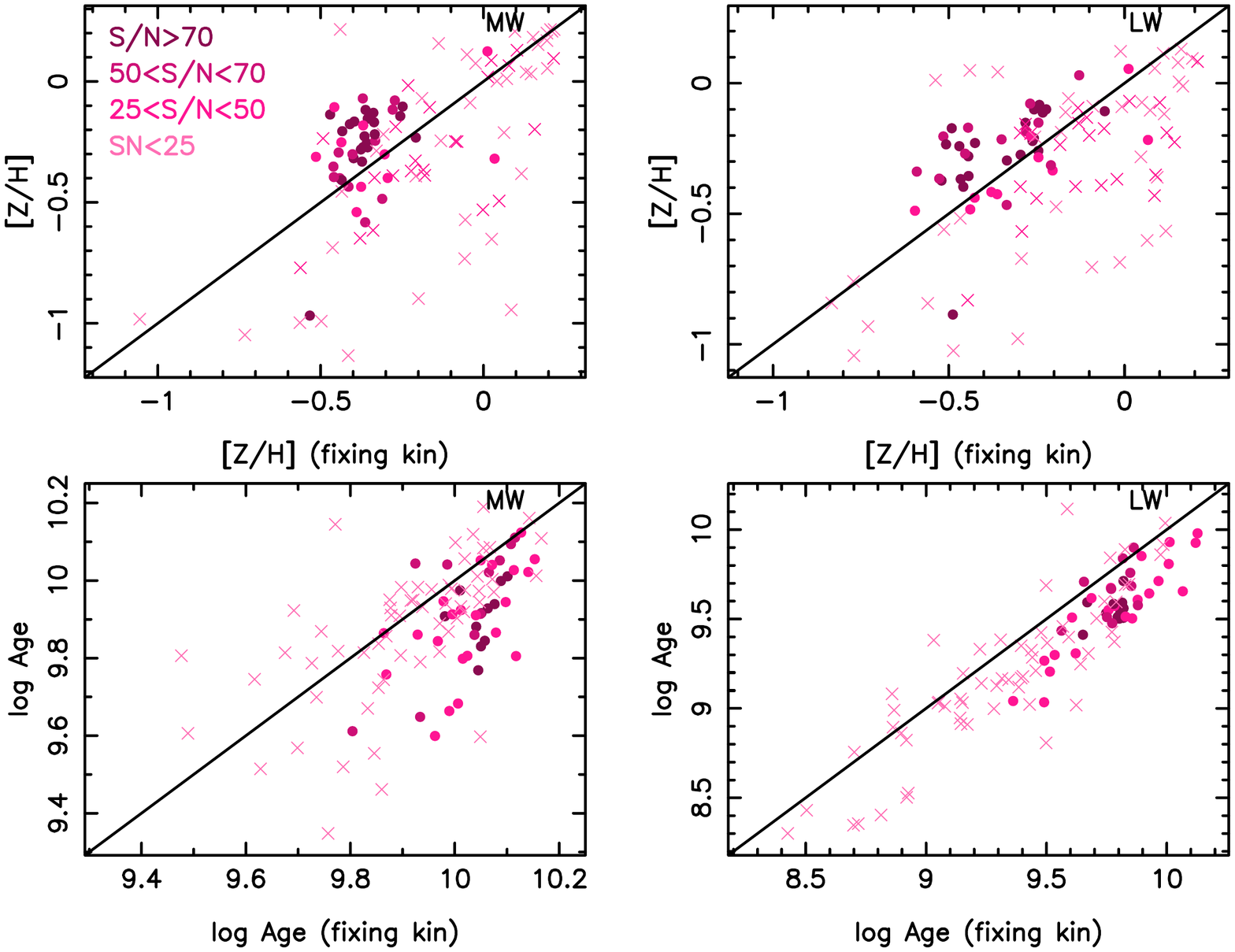}}
\caption{comparison of the mean stellar population parameters with and without fixing the 
kinematics. As can be seen, while the ages are not strongly affected by the choice of 
fixing or not the velocity and velocity dispersion, the metallicities are, due to the 
degeneracy between metallicity and line-broadening. The color of the symbols represent
the S/N of the individual spectrum as indicated in the label.\label{compara_kin}}
\end{figure*}

It is known that the measurement of the velocity dispersion depends 
on the metallicity of the stellar template (Laird \& Levison 1985\nocite{1985AJ.....90.2652L}; 
Bender 1990\nocite{1990A&A...229..441B}; Koleva, Prugniel \& De Rijcke 2008\nocite{2008AN....329..968K}). A deeper absorption feature can be obtained 
by either increasing the metallicity of decreasing the broadening, which leads
to degeneracies when trying to deriving both parameters at the same time. 
To quantify the differences in the derived parameters with and without 
fixing the kinematics, we have run {\tt STECKMAP} allowing to fit both, the radial velocity 
and velocity dispersion of the voronoi-binned spectra in NGC~628 and compare the derived, 
mean parameters, with those obtained when fixing the kinematics (that is the approach followed
in the main paper). The results are shown in Fig.~\ref{compara_kin}. 

As can be seen, while the mean ages are not strongly affected by fixing or not the kinematics, 
the metallicities can be by as much as  $\sim$ 1 dex. The net effect is that, without 
fixing the kinematics, the metallicities tend to be lower . Furthermore, there is a trend
for which the differences increase with the mean metallicity of the population. 
We showed in S\'anchez-Bl\'azquez et al. (2011) that this effect is not exclusive of our
employed technique to derive star formation histories, but that it is also seen if other 
codes, as {\tt starlight} is used.
The differences, also depend on the signal-to-noise. To further emphasize this, we also 
include some spectra resulting from a voronoi-binned spectra imposing a lower signal-to-noise cat
(showed with different symbols in the figure). It can be seen that the lower the signal-to-noise, 
the larger the differences.

\section{Effect of fitting the continuum}
\label{appendix_cont}
\begin{figure*}
\resizebox{0.6\textwidth}{!}{\includegraphics[angle=-90]{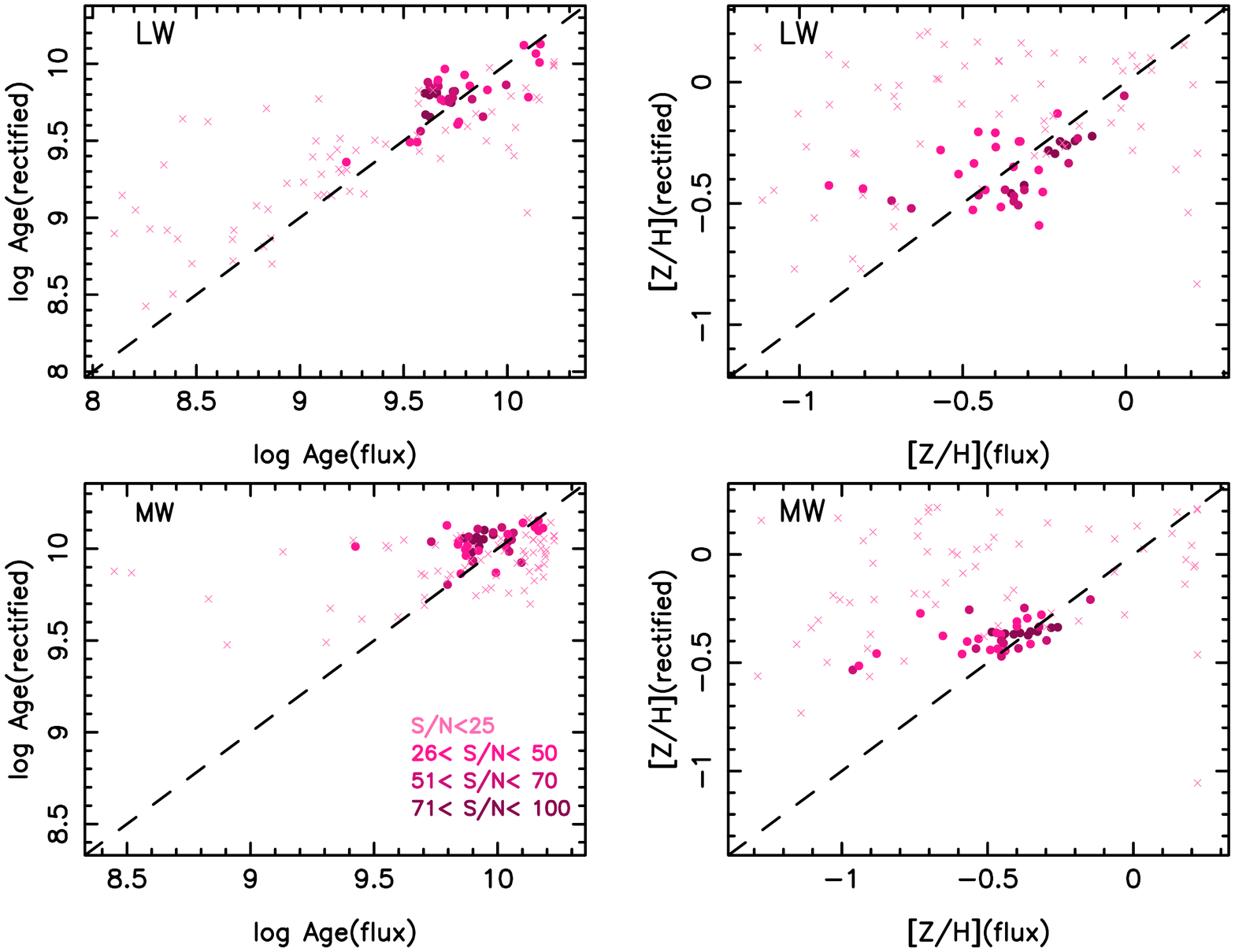}}
\caption{Comparison of the mean stellar population parameters
  obtained when  fitting (flux) and not fitting (rectified) the
  continuum. Darker symbols indicate higher signal-to-noise spectra. The crosses indicate the 
  results for spectra with lower S/N than our minimum cut, included here to illustrate 
  the effect of working with low S/N spectra.\label{fit_cont}}
\end{figure*}

Figure~\ref{fit_cont} shows the comparison of our derived mean stellar  age and metallicity (weighted with both, light
and mass) and that obtained when we allow for the continuum shape to be fit. We have use the spectra of the Voronoi-binned
data. As can be seen in the figures, except for the luminosity-weighted age, the rest of the parameters are strongly 
affected by fitting or not the continuum. It is also clear that for the   those spectra with the highest signal-to-noise (dark
colored symbols), the
differences are much lower than for the spectra with the lowest signal-to-noise (light color and crosses.)

\section{Line-strength Lick indices maps}
\label{appendix_lickmap}
The maps for all the Lick/IDS indices are available
in the electronic edition of the paper. The indices have been measured in the voronoi-binned spectra
broadened to the total resolution of 14\AA (FWHM), that is, matching the LIS-14
spectral system defined in Vazdekis et al.\~(2010).
We measured all the Lick/IDS indices with the definitions of Trager (1998)\nocite{1998ApJS..116....1T} and the 4000-break
with the definition given in Bruzual et al. (1993)\nocite{1993ApJ...405..538B}.

\label{lastpage}

\end{document}